\documentclass[fleqn,usenatbib]{mnras}
\usepackage[T1]{fontenc}
\usepackage{graphicx} 
\usepackage{amsmath}
\usepackage{amssymb}
\usepackage{color}
\usepackage{multicol}
\usepackage{threeparttable}
\usepackage{subcaption}
\usepackage{float}
\usepackage{soul}
\usepackage{bold-extra}
\usepackage{newtxtext,newtxmath}

\DeclareRobustCommand{\VAN}[3]{#2}
\let\VANthebibliography\thebibliography
\def\thebibliography{\DeclareRobustCommand{\VAN}[3]{##3}\VANthebibliography}

\def\refjnl#1{{\rm#1}}
\def\apj{\refjnl{ApJ}}                 
\def\apjl{\refjnl{ApJ}}                
\def\apjs{\refjnl{ApJS}}               
\def\ao{\refjnl{Appl.~Opt.}}           
\def\aap{\refjnl{A\&A}}                
\def\mnras{\refjnl{MNRAS}}             
\def\physrep{\refjnl{Phys.~Rep.}}   

\def\be{\begin{equation}} 
\def\ee{\end{equation}}

\def\msun{{\Msun}}

\def\HI{\hbox{H~$\scriptstyle\rm I\ $}}

\def\gsim{\lower.5ex\hbox{\gtsima}} 
\def\lsim{\lower.5ex\hbox{\ltsima}} \def\gtsima{$\; \buildrel > \over 
\sim \;$} \def\ltsima{$\; \buildrel < \over \sim \;$} \def\prosima{$\; 
\buildrel \propto \over \sim \;$} \def\gsim{\lower.5ex\hbox{\gtsima}} 
\def\lsim{\lower.5ex\hbox{\ltsima}} 
\def\simgt{\lower.5ex\hbox{\gtsima}} 
\def\simlt{\lower.5ex\hbox{\ltsima}} 
\def\simpr{\lower.5ex\hbox{\prosima}} \def\la{\lsim} \def\ga{\gsim} 
  
 \def\gtsima{$\; \buildrel > \over \sim \;$} 
\def\ltsima{$\; \buildrel < \over \sim \;$} 
\def\gsim{\lower.5ex\hbox{\gtsima}} 
\def\lsim{\lower.5ex\hbox{\ltsima}} 
\def\simgt{\lower.5ex\hbox{\gtsima}} 
\def\simlt{\lower.5ex\hbox{\ltsima}} 
\def\simpr{\lower.5ex\hbox{\prosima}} 
\def\la{\lsim} 
\def\ga{\gsim}

\def\msun{\,{\rm \Msun}}

\def\E3{{\cal E}_{\rm g}^{III}}

\def\msun{\rm M_\odot}

\def\M*{M_*}

\def\Z*{Z_*}
\def\L*{L_*}
\def\muv{\rm M_{UV}}

\title[SFHs in the EoR]{Astraeus IV: Quantifying the star formation histories of galaxies in the Epoch of Reionization}

\author[L. Legrand et al.]{Laurent Legrand,$^{1}$\thanks{E-mail: legrand@astro.rug.nl}
Anne Hutter,$^{1}$
Pratika Dayal,$^{1}$
Graziano Ucci,$^{1}$
Stefan Gottl\"{o}ber,$^{2}$
Gustavo Yepes$^{3,4}$
\\
$^{1}$ Kapteyn Astronomical Institute, University of Groningen, P.O Box 800, 97000 AV Groningen, The Netherlands\\
$^{2}$ Leibniz-Institut f\"ur Astrophysik, An der Sternwarte 16, 14482 Potsdam, Germany\\
$^{3}$ Departamento de F\'isica Te\'orica, M\'odulo 8, Facultad de Ciencias, Universidad Autonoma de Madrid, 28049 Madrid, Spain\\
$^{4}$ CIAFF, Facultad de Ciencias, Universidad Autonoma de Madrid, 28049 Madrid, Spain
}

\date{Accepted XXX. Received YYY; in original form ZZZ}

\pubyear{2021}

\begin{document}
\label{firstpage}
\pagerange{\pageref{firstpage}--\pageref{lastpage}}
\maketitle


\begin{abstract}
We use the \textsc{astraeus} framework, that couples an N-body simulation with a semi-analytic model for galaxy formation and a semi-numerical model for reionization, to quantify the star formation histories (SFHs) of galaxies in the first billion years. Exploring four models of radiative feedback, we fit the SFH of each galaxy at $z>5$ as $\mathrm{log}(\mathrm{SFR}(z))=-\alpha(1 + z)+\beta$; star formation is deemed stochastic if it deviates from this fit by more than $\Delta_\mathrm{SFR}=0.6\,$dex. Our key findings are: (i) The fraction of stellar mass formed and time spent in the stochastic phase decrease with increasing stellar mass and redshift $z$. While galaxies with stellar masses of $M_\star\sim10^7\msun$ at $z\sim5~(10)$ form $\sim70\%~(20\%)$ of their stellar mass in the stochastic phase, this reduces to $<10\%$ at all redshifts for galaxies with $M_\star > 10^{10}\msun$; (ii) the fractional mass assembled and lifetime spent in the stochastic phase do not significantly change with the radiative feedback model used; (iii) at all redshifts, $\alpha$ increases (decreases for the strongest radiative feedback model) with stellar mass for galaxies with $M_\star\lsim 10^{8.5}\msun$ and converges to $\sim0.18$ for more massive galaxies; $\beta$ always increases with stellar mass. Our proposed fits can reliably recover the stellar masses and mass-to-light ratios for galaxies with $M_\star\sim10^{8-10.5}\msun$ and $\muv\sim-17~{\rm to}~-23$ at $z\sim 5-9$. This physical model can therefore be used to derive the SFHs for galaxies observed by a number of forthcoming instruments.
\end{abstract}




\section{Introduction}

The earliest galaxies ushered in the Epoch of Reionization (EoR) as their photons (with energies $>13.6$eV) gradually ionized the neutral hydrogen (\HI) gas in the intergalactic medium \citep[IGM; for reviews see e.g][]{barkana-loeb2001, dayal2018}. Understanding this last major phase transition of the Universe therefore naturally requires a detailed picture of the number density, physical properties and large-scale distribution of early galaxies. Next-generation facilities, such as the James Webb Space Telescope (JWST\footnote{https://www.jwst.nasa.gov/}) and the Nancy Grace Roman Space Telescope (NGRST\footnote{https://www.nasa.gov/content/goddard/about-nancy-grace-roman-space-telescope}), aim at measuring the ultra-violet (UV) light emitted by these first galaxies in unprecedented detail. However, deriving the corresponding stellar populations and galactic properties (such as stellar masses, star formation rates or mass-weighted stellar ages) from the measured spectral energy densities (SEDs) will be complicated requiring, amongst other parameters, a detailed understanding of their star formation histories  \citep[SFHs;][]{lower2020measure}. For instance, the SFHs recovered from SEDs of $z\simeq1-5$ galaxies can vary from exponentially rising to exponentially declining, with the uncertainty in the redshift trend increasing with increasing redshift \citep{Ciesla2017}.

Observationally, the SFHs of galaxies can be derived from their stellar populations. While this approach is feasible for galaxies in the local Universe \citep[e.g.][]{Gallart_2015, Albers_2019}, direct observations of the stellar populations of galaxies at higher redshifts is quite impossible. However, analysing the SEDs of high-redshift galaxies that contain the integrated light from their stellar populations and their surrounding ionized gas represents an indirect and alternative method of constraining the recent SFH \citep[$\lesssim100$~Myr;][]{Calzetti2013}. Here, while the H$\alpha$ recombination line \citep{Kennicutt_1998} traces the most recent star formation (within the last $\sim10\,\mathrm{Myr}$), the far-ultraviolet continuum \citep{Kennicutt_2012} is sensitive to star formation within the last $\sim100\,\mathrm{Myr}$. Nevertheless, the galactic properties derived by this method still depend on the assumed shape of the past SFH (at ages $\gtrsim100$~Myr). Moreover, the derived SFH is highly sensitive to small variations in the SED data \citep{Ocvirk2006}. 
Given the absence of further observational constraints on the SFH at higher redshifts, a large range of SFH models have been employed in different works, ranging from (possibly delayed) exponentially declining \citep{Ciesla_2015, Wilkinson2017} and constant SFHs \citep{Yoon2021} to log-normal \citep{Diemer_2017} and double power laws \citep{Behroozi2013, Carnall2018} for galaxies at $z\simeq0-1$.  For high-redshift galaxies during the EoR, the SFHs used to interpret observations range from exponentially declining \citep[e.g.][]{Robert-Borsani2021} to increasing as a power-law with decreasing redshift \citep[e.g.][]{Song_2016}.

The high uncertainties in extracting the SFHs of galaxies from observations alone therefore require theoretical inputs to understand the key physical processes governing the SFH. On the one hand, the SFH of a galaxy is determined by the amount of gas gained through mergers and accretion that replenishes the cold gas reservoir and fuel star formation. On the other hand, feedback processes can reduce the cold gas content and suppress subsequent star formation. Key feedback processes during the EoR include the heating and ejection of gas through supernova (SN) explosions \citep{MacLow1999}, and radiative feedback from reionization, i.e. the photo-evaporation of gas \citep{Barkana_1999, Shapiro_2004, Iliev_2005} or the suppression of gas infall \citep{Gnedin2000, Hoeft_2006} due to the photo-heating of gas in ionized regions. The efficiency with which both these feedback mechanisms suppress star formation decreases as the gravitational potential of a galaxy deepens. Hence, while low-mass galaxies are expected to exhibit episodes of star formation upon gas accretion or gas-rich mergers followed by episodes of significant/complete suppression of star formation, also referred to as ``stochastic star formation", star formation in massive galaxies is less susceptible to SN and radiative feedback, and hence more ``continuous" in nature. 

Hierarchical structure formation predicts that galaxies grow both in dark matter (DM) and gas mass over time. This suggests that star formation in early galaxies increases with time during the EoR, transitioning from being stochastic to continuous \citep[see e.g.][]{dayal2013, Kimm2014, Faisst2019, Emami2019} as their gravitational potentials become deep enough to withstand SN and radiative feedback. Indeed, hydrodynamical simulations assuming a homogeneous photoionization background at $z>6$ have found that the average SFH is smoothly rising with time and scales with stellar mass, and that the SFH shape is scale-invariant \citep{Finlator_2010}. The rising trend of the average SFH has also been confirmed by radiative hydrodynamical simulations of the EoR that account for feedback from both SN and an inhomogeneous UV background (UVB) for galaxies with halo masses of $M_\mathrm{h}\gtrsim10^9\msun$ \citep{Ocvirk2016cosmic, Ocvirk2020}. In principle, these SFHs are in agreement with the individual SFHs of $z=0$ galaxies in the Illustris simulation. \citet{Diemer_2017} found these SFHs to be characterised by a log-normal function in time, which implies a rising SFH at early times for most galaxies. While traditionally SFHs have been quantified by finding accurate fitting functions with least square methods \citep[e.g.][]{Diemer_2017} or Bayesian interference \citep[e.g.][]{Carnall2018}, recent approaches have begun to employ machine learning that extract a direct relation between the spectrum of a galaxy and its SFH \citep[see e.g.][]{Lovell2019}.

While radiative hydrodynamical simulations can follow the time and spatial evolution of the ionization of the IGM and the physical processes in early galaxies (including gas accretion, stellar and radiative feedback processes), they remain computationally very expensive. This has limited investigations to either running a single cosmological simulation (volumes $\lesssim10^6$~comoving~Mpc$^3$) following galaxy evolution and reionization \citep[see e.g.][]{Ocvirk2020} or exploring different physical process descriptions in smaller volumes \citep[e.g.][]{Gnedin2014, Yajima2017, Smith2019}.
However, semi-numerical models of galaxy evolution and reionization using dark matter-only N-body simulations as inputs for the mass assembly of galaxies and IGM matter distribution \citep{Iliev2006, Mutch2016, Dixon2018, Seiler2019, Hutter2021astraeusI} require significantly less computational resources and provide the ideal tool to explore the implications of different galactic and intergalactic physical processes on the SFHs and galactic properties for a representative galaxy population (i.e. covering volumes $>10^6$~cMpc$^3$). Such models not only resolve the low-mass galaxies that might have driven reionization, but also track the evolution and large-scale distribution of the associated ionized regions \citep[e.g.][]{Hutter2021ast3}. Moreover, describing galaxy evolution and reionization self-consistently, they allow us to explore the processes that shape the interplay between galaxy evolution and reionization and hence the SFHs, such as radiative feedback or the escaping ionizing emissivity from early galaxies. 

In this paper, we will use the \textsc{astraeus} framework (\citealt{Hutter2021astraeusI}), which couples the dark matter merger trees and matter density distributions obtained from an N-body simulation with a semi-analytic galaxy evolution model and a self-consistent semi-numerical reionization scheme, to quantify the SFHs of galaxies during the EoR and assess how strongly their SFHs are shaped by radiative feedback. In particular, we pursue the following questions: What is the functional form of the SFHs of galaxies during the EoR and how do they evolve with redshift and stellar mass? How much stellar mass is assembled during the initial period of stochastic star formation? How do the SFHs depend on the strength of the radiative feedback model used?
The \textsc{astraeus} model is ideally suited for such an investigation: Firstly, it supports multiple models for radiative feedback and the ionizing escape fraction which allows us to cover the plausible parameter space of galaxy evolution and reionization scenarios. And secondly, the underlying N-body simulation comprises a large enough volume and high mass resolution to simulate a representative galaxy population.

\begin{center}
\begin{table*}
\caption{\label{tab:models} For the radiative feedback model shown in column 1, we show the threshold star formation efficiency (column 2), the fraction of kinetic SNII energy deposited in the ISM (column 3), the escape fraction of hydrogen ionizing photons (column 4), the characteristic mass (column 5) and the temperature of the ionized IGM (column 6).}

\begin{tabular}{|c | c c c c c |}
\hline
Model & $f_*$ & $f_\mathrm{w}$ & $f_\mathrm{esc}$ & $M_\mathrm{c}$ & $T_\mathrm{0}$\\ [0.5ex]
\hline\hline   
\textit{Photoionization} & 0.01 & 0.2 & 0.215 & $M_\mathrm{c}(z, z_{reion}, \Gamma_{HI})$ & a\footnotemark[1] \\
\hline 
\textit{Early Heating} & 0.01 & 0.2 & 0.60 & $M_\mathrm{F}(z, z_{reion}, T)$ & $4\times 10^4\,\rm K$\\
\hline
\textit{Strong Heating} & 0.011 & 0.19 & 0.22 & $8M_\mathrm{F}(z, z_{reion}, T)$ & $4\times 10^4\,\rm K$\\
\hline
\textit{Jeans Mass} & 0.01 & 0.2 & 0.285 & $M_\mathrm{J}(z, T)$ & $4\times 10^4\,\rm K$\\ [1ex]
\hline
\end{tabular}
\begin{tablenotes} \scriptsize
\item[a$^1$] $T_0$ is set by the photoionization rate at the moment a galaxy's environment is reionized
\end{tablenotes}

\end{table*}
\end{center}

This paper is structured as follows. In Sec.~\ref{Model}, we briefly describe the \textsc{astraeus} framework, including the underlying N-body simulation, the different physical processes implemented in the galaxy evolution and reionization model and the radiative feedback models investigated. We summarise the relevant physical processes that determine the star formation in early galaxies. In Sec.~\ref{Results}, we show the SFHs obtained, describe the assumed functional form for the SFHs and our fitting procedure and quantify how the transition from stochastic to continuous star formation evolves with redshift and stellar mass. Finally, in Sec.~\ref{Validation}, we confirm that the derived fitting functions reproduce the stellar masses and UV luminosities of the underlying galaxy population before concluding in Sec.~\ref{Conclusions}.

\section{The theoretical model} \label{Model}

We use the \textsc{astraeus} (semi-numerical rAdiation tranSfer coupling of galaxy formaTion and Reionization in N-body dArk mattEr simUlationS) framework that couples a state-of-the-art N-body simulation run as part of the Multi-dark project\footnote{www.cosmosim.org} (\textit{Very small multi-dark Planck}; \textsc{vsmdpl}) with a slightly modified version of the \textsc{delphi} semi-analytic model of galaxy formation \citep{Dayal2014} and the \textsc{cifog} (Code to compute ionization field from density fields and source catalogue) semi-numerical reionization scheme \citep{Hutter_2018}. We briefly describe the model here and interested readers are referred to \citet{Hutter2021astraeusI} for complete details. \\

The underlying dark matter only N-body simulation\footnote{The following cosmological parameters are assumed: [$\rm \Omega_\Lambda$, $\rm \Omega_m$, $\rm \Omega_b$, $\rm h$, $\rm n_s$, $\rm \sigma_8$] = [0.69, 0.31, 0.048, 0.68, 0.96, 0.82].} has been run using the \textsc{gadget-2} Tree+PM code \citep{Springel_2005, Klypin_2016}. It has a box side length of $160h^{-1}$Mpc and follows $3840^3$ particles, with each particle having a dark matter mass of $m_\mathrm{DM}=6.2\times10^6h^{-1}\msun$. We use the phase space halo finder \textsc{rockstar} (\citealt{Behroozi2012a}) to identify dark matter halos and the \textsc{consistent trees} algorithm (\citealt{Behroozi2012b}) to derive merger trees which have been resorted to local horizontal (sorted on a redshift-by-redshift-basis within a tree) merger trees using the \textsc{cutnresort} module within the \textsc{astraeus} pipeline (see Appendix A in \citealt{Hutter2021astraeusI}). For all snapshots, the dark matter density fields have been re-sampled to a $512^3$ grid which are used as input files for the \textsc{astraeus} code. From the 150 snapshots from $z=25$ to $z=0$, we employ the first 74 snapshots, ranging from $z=25$ down to $z=4.5$. Although the properties of galaxies in \textsc{astraeus} converge for halos down to $50$ dark matter particles \citep[see Appendix B in][]{Hutter2021astraeusI}, we limit all analyses presented in this work to halos with at least $100$ particles, corresponding to $M_h = 6.2\times 10^8\,h^{-1}\mathrm{M_\odot}$; this ensures the convergence of their SFHs. Finally, we use a Salpeter initial mass function \citep[IMF;][]{salpeter1955} between $0.1-100\msun$ throughout this work.

\begin{figure*}
    \centering
    \includegraphics[width=\textwidth]{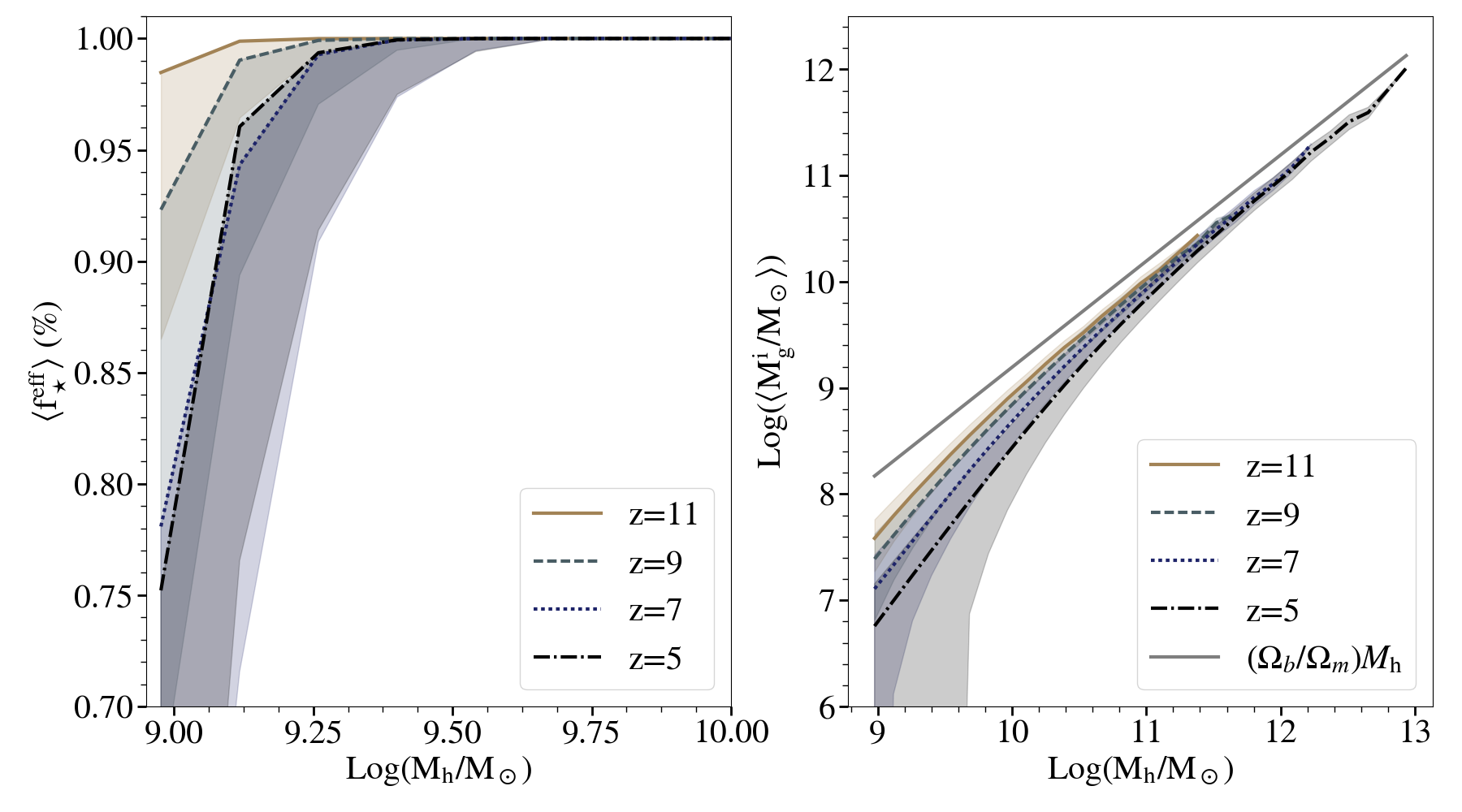}
    \caption{Mean effective star formation efficiency ($\langle f\mathrm{_\star^{eff}}\rangle$, \textit{left panel}) and mean initial gas mass ($\langle M_\mathrm{g}^\mathrm{i} \rangle$, \textit{right panel}) as a function of the halo mass. In both panels, the lines show results for the redshifts marked, with the shaded regions showing the $1-\sigma$ standard deviation. In the \textit{right panel}, the black solid line shows the gas mass-halo mass relation assuming a cosmological baryon-to-matter ratio.}
    \label{fig:feffMgasIni}
\end{figure*}

The N-body simulation is coupled to a modified version of the semi-analytic galaxy evolution code \textsc{delphi} (\citealt{Dayal2014}) which tracks the joint evolution of dark matter halos and their baryonic components accounting for all key physical  processes including mergers (of dark matter, gas and stellar mass), smooth accretion of dark matter and gas from the IGM, and feedback (from both SN and reionization). In brief, a newly-formed halo (a ``starting leaf") at redshift $z$ with mass $M_\mathrm{h}(z)$ is assigned a gas mass of $M_\mathrm{g}^\mathrm{i}(z) = (\Omega_b / \Omega_m)M_\mathrm{h}(z)$, where $\Omega_\mathrm{b}/\Omega_\mathrm{m}$ is the cosmological baryon-to-matter ratio. However, a halo that has progenitors can gain gas through: (i) mergers where the merged gas mass is $M_\mathrm{g}^\mathrm{mer}(z) = \sum_p M_\mathrm{g, p}(z+\Delta z)$. Here, the RHS denotes the final gas mass brought in by all progenitors from the previous redshift step; and (ii) smooth accretion from the IGM where accretion of a dark matter mass  $M_\mathrm{h}^\mathrm{acc}(z)$ is assumed to be accompanied by a cosmological ratio of gas mass such that $M_\mathrm{g}^\mathrm{acc}(z) = (\Omega_\mathrm{b} / \Omega_\mathrm{m})M_\mathrm{h}^\mathrm{acc}(z)$. However, galaxies in ionized regions can lose all or part of their initial gas mass due to photo-heating by reionization radiative feedback such that the initial gas mass is
\begin{eqnarray}
    M_\mathrm{g}^\mathrm{i}(z) &=& \mathrm{min} \left[ M_\mathrm{g}^\mathrm{acc}(z) + M_\mathrm{g}^\mathrm{mer}(z), f_\mathrm{g} \frac{\Omega_b}{\Omega_m}M_\mathrm{h}(z) \right],
    \label{eq:mgas}
\end{eqnarray}
where $f_\mathrm{g}$ is the fraction of gas still available after radiative feedback.
A fraction $f_\star^\mathrm{eff}$ of this initial gas mass is turned into stars at each redshift-step. This ``effective efficiency" is the minimum between that required to eject the rest of the gas and quench star formation ($f_\star^\mathrm{ej}$) and an upper limit ($f_\star$), such that $f_\star^\mathrm{eff}=min[f_\star^\mathrm{ej}, f_\star]$. We account for mass-dependent stellar lifetimes \citep{padovani1993} in order to calculate the stars that explode as TypeII SN (SNII) within a given redshift-step. 
In this formalism, $f_\star^\mathrm{ej}$ can be calculated as
\begin{equation}
    f_\star^{ej}(z) = \frac{M_\star^\mathrm{new}(z)}{M_\mathrm{g}^\mathrm{ej}(z) + M_\star^\mathrm{new}(z)} ,
   \label{eq:feff}
\end{equation}
where $M_\star^\mathrm{new}(z)$ and $M_\mathrm{g}^\mathrm{ej}(z)$ are the newly formed stellar mass and the ejected gas mass at redshift $z$, respectively.  

\vspace{0.2cm}

Reionization is included in our framework through the \textsc{cifog} semi-numerical scheme  (\citealt{Hutter_2018}) which computes the time and spatial evolution of hydrogen ionization fields. At each redshift-step, the number of ionizing photons produced by each galaxy is calculated using the stellar population synthesis code \textsc{starburst99} (\citealt{Leitherer_1999}); the inputs for this include the entire SFH, the IMF and our assumption of a stellar metallicity $Z = 0.05 \,Z_\odot$. However, only a fraction of these ionizing photons, $f_\mathrm{esc}$, can escape the galaxy and contribute to the reionization of the IGM. If in a region, the cumulative number of ionizing photons emitted exceeds the cumulative number of absorption events, this region is considered ionized and the temperature rises to $T_0$; otherwise the region is considered neutral. Galaxies in ionized regions are then subject to radiative feedback. We explore four different radiative feedback scenarios in this work that are characterized by different prescriptions for the characteristic mass ($M_\mathrm{c}$) - this corresponds to the halo mass at which a galaxy can retain half of its gas compared to the cosmological ratio as now detailed:

\begin{itemize}
     \item \textbf{\textit{Photoionization}}: In this model, $M_c$ is given by the fitting function derived from 1D radiation-hydrodynamical simulations \citep{Sobacchi_2013}. $M_c$ increases with an increase in the photoionization rate and/or the difference between the reionization redshift and the current galaxy redshift. The escape fraction is assumed to be constant with $f_{esc} = 0.215$. This model results in a \textbf{time delayed, weak} radiative feedback.
     
     \item \textbf{\textit{Early Heating}}: Using simulations of cosmological reionization, \citealt{Naoz_2013} have shown that the characteristic mass is related to the filtering mass $M_\mathrm{F}$ as $M_\mathrm{c}= 1/8 M_\mathrm{F}$. We assume that the ionized IGM has a temperature of $T_0=4\times 10^4\,\rm K$. Additionally, in contrast to the other models, $f_\mathrm{esc}$ depends on the fraction of gas ejected from the galaxy, such that $f_\mathrm{esc} = f_0 (f_\star^\mathrm{eff}/f^\mathrm{ej}_\star)$; $f_0=0.6$ is a free parameter that is tuned to reproduce the reionization optical depth. This model results in a \textbf{time delayed, weak to intermediate} radiative feedback.
     
     \item \textbf{\textit{Strong Heating}}:  In this model, the IGM is also heated to $T_0=4\times 10^4\,\rm K$ upon reionization but the characteristic mass equals the filtering mass ($M_c=M_F$), allowing us to explore the effect of stronger radiative feedback. For all galaxies and redshifts, the escape fraction in this model is assumed to be constant such that $f_{esc} = 0.22$. This model results in a \textbf{time delayed, maximum} radiative feedback.
    
     \item \textbf{\textit{Jeans Mass}}: In this model, the gas density is assumed to react instantaneously to the gas temperature increasing to $T_0=4\times 10^4\,\rm K$ in ionized regions. Consequently, $M_\mathrm{c}$ equals the Jeans mass $M_\mathrm{J}(z)$ at the virial over-density as soon as a galaxy's environment becomes reionized. For all galaxies and redshifts, the escape fraction in this model is assumed to be constant with $f_{esc} = 0.285$. Hence, the \textit{Jeans Mass} model results in an \textbf{instantaneous, maximum} radiative feedback. 
\end{itemize}

The model therefore contains three redshift-independent free parameters: $f_\star$, $f_\mathrm{w}$ and $f_\mathrm{esc}$. The first two, which are also mass-independent, are tuned to reproduce key \textit{galaxy observables}, such as the evolving UV luminosity function (UVLF), the stellar mass function (SMF) and the cosmic star formation rate density (SFRD) at $z=5-10$. We tune $f_\mathrm{esc}$ to reproduce \textit{reionization observables} including the Thomson scattering optical depth and constraints on the reionization history inferred from quasars, Lyman Alpha Emitters and Gamma-Ray Bursts. The models used in this work are summarised in \textbf{Table~\ref{tab:models}}.

\begin{figure*}
\centering
   \includegraphics[width=\textwidth]{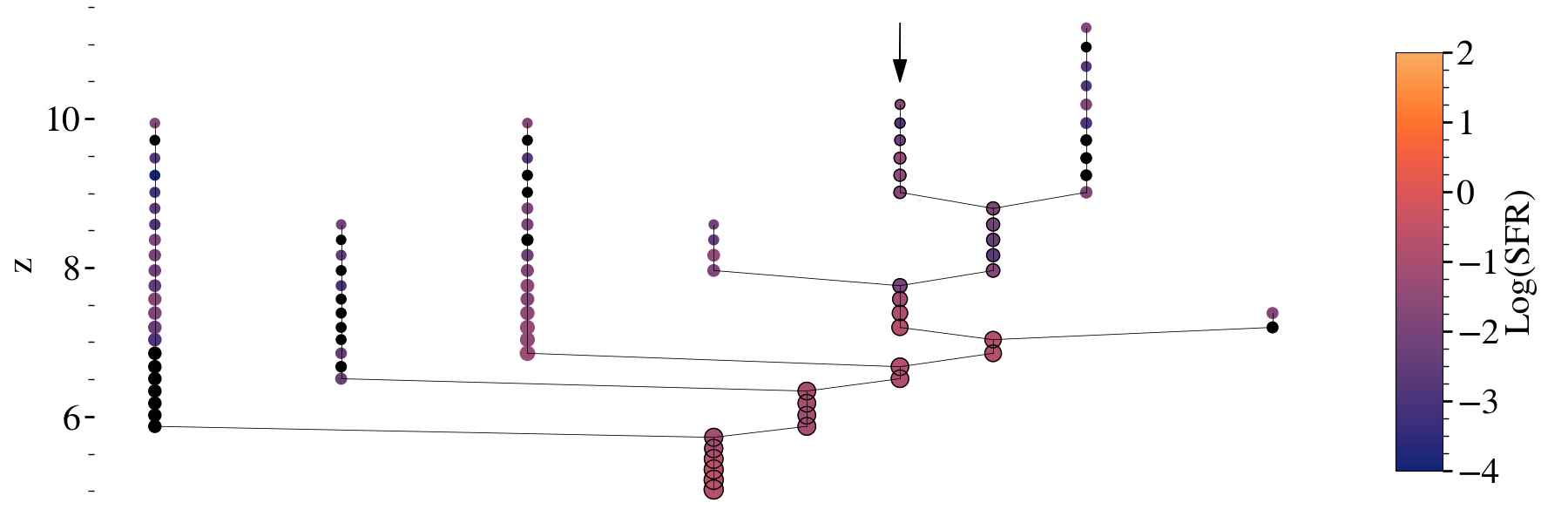}
   \includegraphics[width=\textwidth]{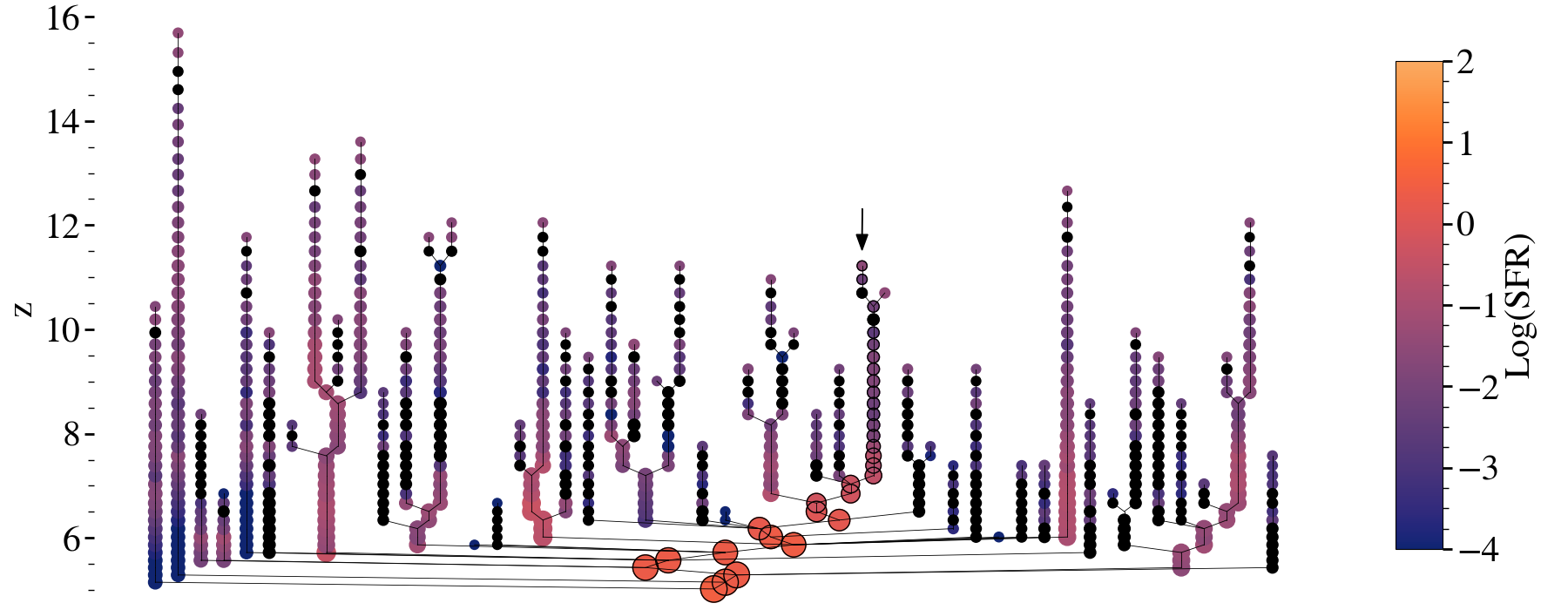}
   \includegraphics[width=\textwidth]{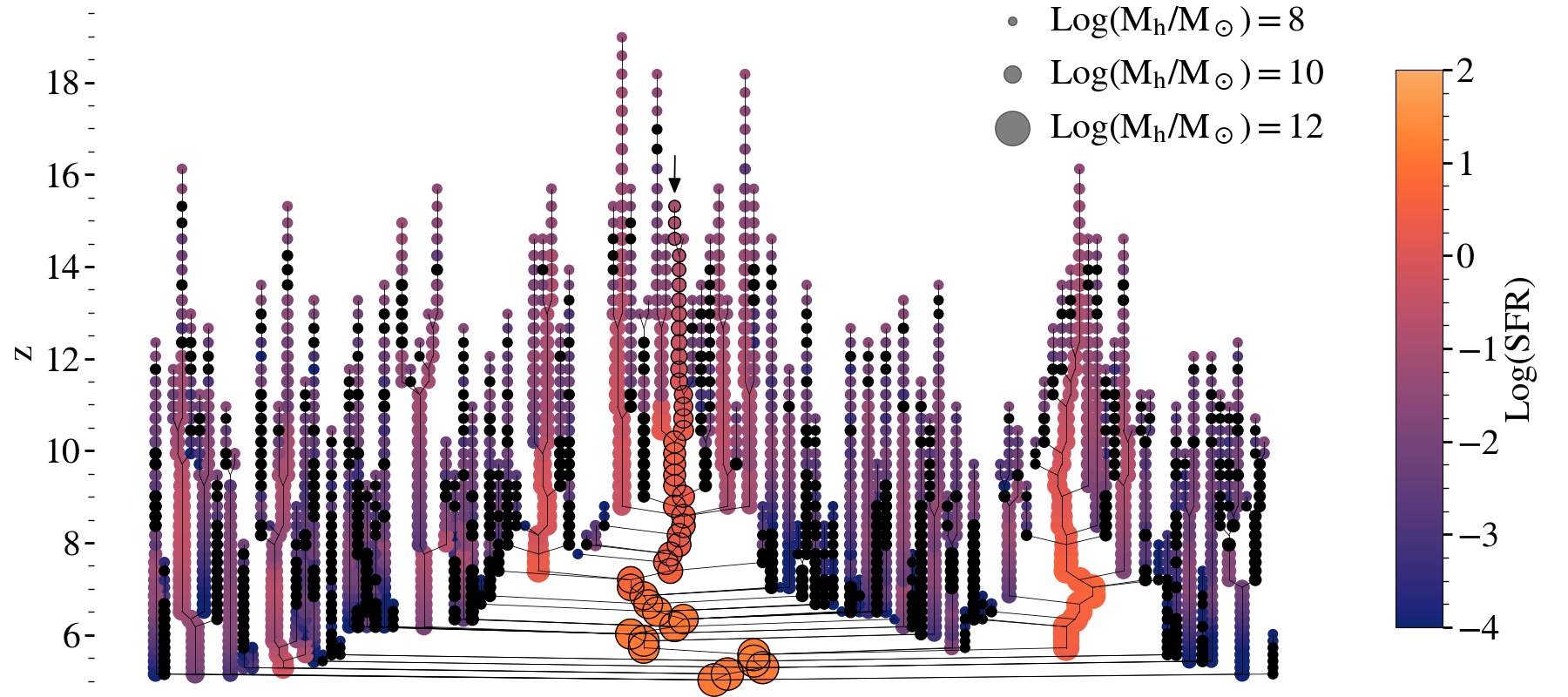}
    \caption{The merger trees for a low-mass galaxy ($M_\star = 10^8\,\msun$, $M_\mathrm{h} = 10^{10.3}\,\msun$, top panel), an intermediate-mass galaxy ($M_\star = 10^9\,\msun$, $M_\mathrm{h} = 10^{11.2}\,\msun$, middle panel) and a massive galaxy ($M_\star = 10^{10}\,\msun$, $M_\mathrm{h} = 10^{11.8}\,\msun$, bottom panel) at $z=5$. Each progenitor is represented by a filled circle with the color scaling with its star formation rate as per the color bar (black represents the absence of star formation). The size of each circles scales with the halo mass as per the indicative sizes shown. Progenitors encircled by a black line indicate the major branch with the black arrow indicating the starting leaf of the major branch.}
    \label{fig:display_merger}
\end{figure*}

\begin{figure*}
    \centering
    \includegraphics[width=\textwidth]{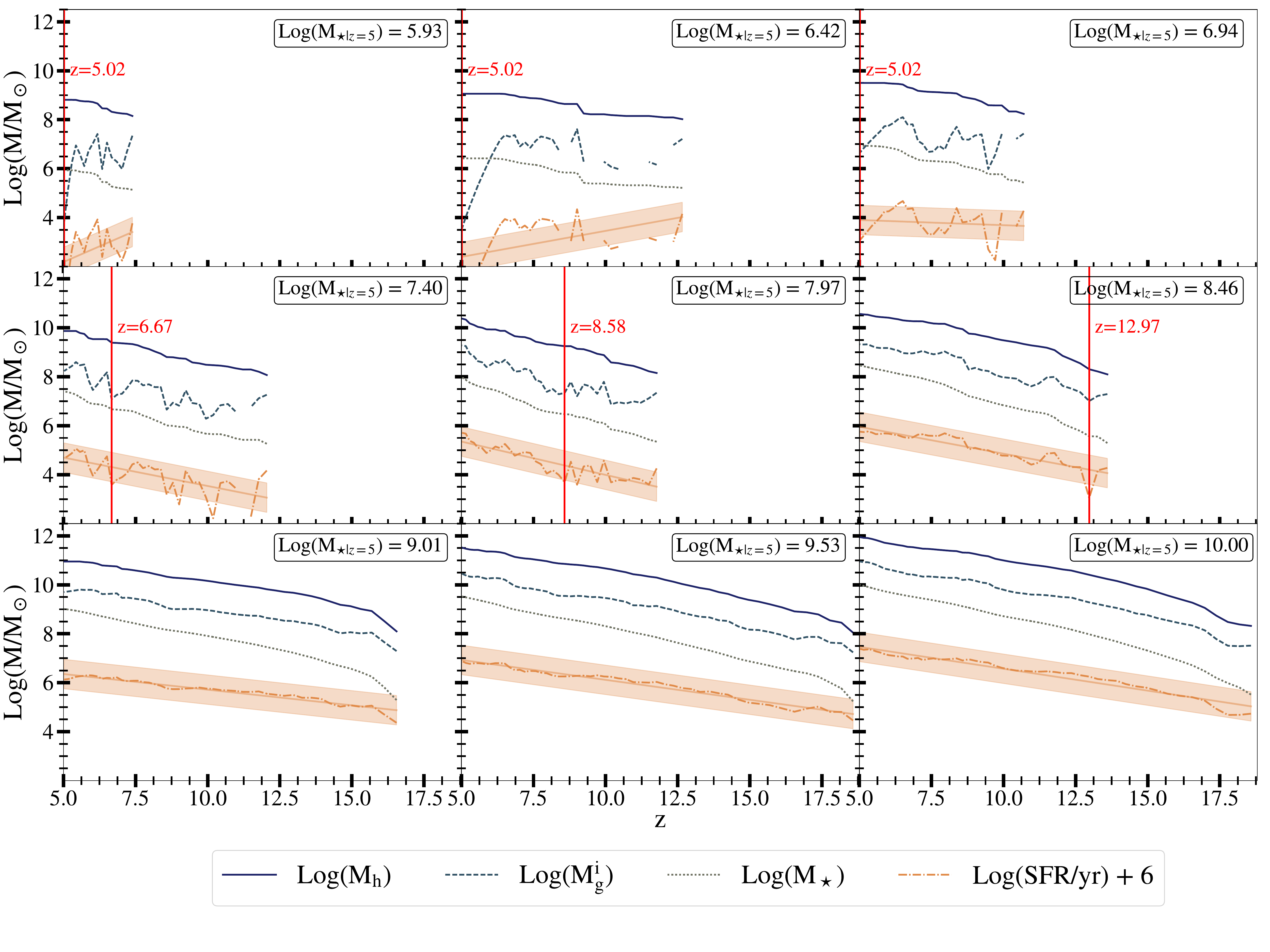}
    \caption{Galaxy assembly as a function of redshift in a model with SNII feedback only. For 9 galaxies of increasing stellar mass at $z \sim 5$, as marked in each panel, we show the evolution of the halo mass ($M_\mathrm{h}$; solid line), the initial gas mass available for star formation ($M_\mathrm{g}^\mathrm{i}$; dashed line), the stellar mass ($M_\star$; dotted line) and the star formation rate (SFR) in units of $\mathrm{Log}(M\mathrm{/M_\odot/yr})$ (dot-dashed line), with the quantity shown being summed over all progenitors at the previous redshift. To illustrate our SFH fitting approach, we also plot the fit of the SFR as the orange solid line and the allowed deviation $\Delta_\mathrm{SFR}$ within which the galaxy is considered non-stochastic as the shaded area. The vertical red line shows the redshift at which each galaxy transitions from a stochastic to a continuous/steady star forming phase.}
    \label{fig:6gal}
\end{figure*}

\subsection{The physical processes determining the star formation rate (SFR)} \label{processes}
At each redshift step, the star formation rate is determined by two interlinked properties: the initial gas mass ($M\mathrm{^i_g}$) and the effective star formation efficiency ($f_\star^\mathrm{eff}$) at which this gas can form stars. Both of these depend on the gravitational potential, the redshift of the halo and the (SN and radiative feedback affected) gas assembly history of a galaxy. 

We start by discussing the evolution of the mean effective star formation efficiency as a function of the halo mass, in a scenario without radiative feedback, as shown in the left panel of Fig.~\ref{fig:feffMgasIni}. Focusing first on $z=5$, we see that galaxies with $M_\mathrm{h} \sim 10^9\,\mathrm{M_\odot}$ have a low value of $\langle f_\star^\mathrm{eff} \rangle \sim 0.75\%$ due to their shallow gravitational potentials; a few SNII are able to push out the remaining gas from such galaxies, quenching subsequent star formation, at least temporarily. These galaxies are in the \textit{feedback limited} phase. The value of $\langle f_\star^\mathrm{eff} \rangle$ increases with $M_\mathrm{h}$ as the gravitational potential deepens. At a transition mass of $M_\mathrm{h} \sim 10^{9.3}\,\mathrm{M_\odot}$, galaxies have a deep enough potential well such that they can form stars at the chosen threshold of $f_\star=1\%$: these galaxies are in the \textit{star formation efficiency limited} phase. The transition mass between the two phases decreases with increasing redshift because galaxies with a given $M_\mathrm{h}$ can support a higher star formation efficiency with increasing redshift due to their deeper potentials (\citealt{Dayal2014}). The scatter in this quantity can be explained as follows: at the low-mass end, $f_\star^\mathrm{eff} = f_\star^{ej}$ with $f_\star^{ej} \propto M_\star^\mathrm{new}$ (see Eqn. \ref{eq:feff}). The value of $M_\star^\mathrm{new}$ can be as low as zero for low-mass halos that have no gas, inducing the scatter in this relation. As expected, high-mass galaxies, which are less feedback affected, show a smaller scatter. 

Next, we discuss the average initial gas mass $\langle M_\mathrm{g}^\mathrm{i} \rangle$ available for star formation as a function of halo mass at $z=5-11$ considering SN feedback only as shown in the right panel of the same figure. We see that at every redshift, $\langle M_\mathrm{g}^\mathrm{i} \rangle$ increases with increasing $M_\mathrm{h}$ given their deeper gravitational potentials. Further, $\langle M_\mathrm{g}^\mathrm{i} \rangle$ approaches the cosmological gas fraction as the halo mass increases. Indeed, as noted above, low-mass galaxies with $M_\mathrm{h} \lsim 10^{9}\, (10^{9.3})\,\mathrm{M_\odot}$ are completely SN-feedback suppressed at $z \sim 5\, (10)$. This leads to a decrease in the gas content of the successor galaxies that they evolve into. Further, given their fewer generations of SN-feedback suppressed progenitors, halos of a given mass show higher gas-to-halo mass ratios with increasing redshift. For example, from this figure we see that $M_h \sim 10^{10}\msun$ halos show a gas mass of $\langle M_\mathrm{g}^\mathrm{i} \rangle \sim 10^{8.2}\msun$ at $z \sim 5$ that increases to $10^{9}\msun$ by $z \sim 11$.

\subsection{The star formation histories of early galaxies} \label{starform}
In order to display the joint evolution of the dark matter halos and their baryonic components, we show the merger trees of a low-mass ($M_\star = 10^8 M_\odot$), an intermediate-mass ($M_\star = 10^9 M_\odot$) and a massive ($M_\star = 10^{10} M_\odot$) galaxy at $z=5$ in Fig.~\ref{fig:display_merger}. Firstly, we note that massive galaxies undergo more mergers throughout their life than intermediate- or low-mass galaxies. For example, when accounting only for mergers in the main branch, the massive $10^{10}\,\msun$ galaxy undergoes $65$ merger events, while the $10^8\,\msun$ and $10^9\,\msun$ galaxies undergo $7$ and $29$ mergers, respectively. Secondly, while the minor branches (those that do not merge directly into the main branch) in low-mass galaxies contain hardly any merger events in their relatively short lifetimes (e.g. the shown mass assembly history of the $M_\star=10^8\msun$ galaxy extends up to $z\simeq11$), minor branches in massive galaxies undergo multiple merger events but have also longer mass assembly histories (up to $z\simeq19$ for the shown $M_\star=10^{10}\msun$ galaxy). Thirdly and most importantly, for all galaxies, irrespective of their final mass, the SFHs of their low-mass progenitors show a large variation in their SFRs including phases of no star formation (black points in merger trees in Fig. \ref{fig:display_merger}). These SFR variations are characteristic of stochastic star formation as defined in Sec.~\ref{characterizing} that follows.

We show the assembly of galaxies with $M_\star \sim 10^{6-10}\,\mathrm{M_\odot}$ at $z=5$ (in a scenario without radiative feedback) in Fig.~\ref{fig:6gal}. We note that the stellar mass and SFHs of a galaxy are summed over all its progenitors at any redshift and that the redshift steps in our simulation increase from $\sim 3\,\mathrm{Myr}$ at $z=25$ to $\sim 37\,\mathrm{Myr}$ at $z=5$. The solid orange line is a linear regression fit to the SFH and the vertical line shows the redshift at which a galaxy transitions from being a ``stochastic" to a ``steady" star former.

As expected, the more massive a galaxy, the earlier it starts assembling. Further, the halo mass increases through accretion and mergers with decreasing redshift, as $\mathrm{Log}(M_\mathrm{h}) \propto -(0.2\sim0.25)z$. Galaxies with $M_\star \sim 10^{6.06-6.93}\,\msun$ (shown in the first row) form in low-mass halos ($M_h \lsim 10^{9.5}\msun$) that are SN feedback limited for most of their assembly history, as explained in Sec.~\ref{processes}. This leads to a highly stochastic assembly of gas mass which is reflected in the burstiness of their SFRs. For example, the galaxy with $M_\star \sim 10^{6.4}\,\mathrm{M_\odot}$ shows multiple episodes where gas has been accreted along with the dark matter and then pushed out of the galaxy by SN feedback between $z \sim 9-13$. Further, the SFH of these low-mass galaxies varies by more than 0.8\,dex around the linear regression fit throughout their history i.e. they are always in the stochastic star formation phase. 
As we go to higher masses and consider the galaxies in the second row in the same figure ($M_\star = 10^{7.40 - 8.51}\,\mathrm{M_\odot}$), on average, the SN feedback following a star formation event in these galaxies expels less than 50\% of their initial gas mass which allows them to sustain further star formation, although at a lower rate. In this case, while the SFH still displays a stochastic behavior, the variations around the fit are smaller, being of the order of 0.6\,dex. These galaxies transition into the steady star forming phase at $z \sim 6.9-8.8$.

Finally, the deep halo potentials ($M_h \sim 10^{11-12}\msun$) of the most massive galaxies (last row of the same figure) with $M_\star = 10^{9.01-10}\,\mathrm{M_\odot}$ ensure that they retain most of their gas. In this case the gas mass and hence the SFR scale with the halo mass at effectively all $z \sim 5-15$, i.e. these galaxies are \textit{always} in the steady star formation phase. The reason for this lacking stochasticity in the early history of massive galaxies is two-fold: Firstly, since galaxies of a given mass have deeper gravitational potentials at higher redshifts, the main branches of their merger tree escape the stochastic phase earlier in their mass assembly histories and correspondingly at lower stellar masses. Secondly, massive galaxies have considerably more progenitors than low-mass galaxies (see Fig. \ref{fig:display_merger}). Hence, as the total SFH of a massive galaxy is constructed by summing up the SFHs of all its progenitors, the stochastic star formation of the low-mass progenitors averages out.

Additionally, the lifetime of a galaxy ($t^\mathrm{tot}$), defined as the time between the current redshift and the redshift of its first progenitor, increases with stellar mass, as expected in hierarchical structure formation. For example, while the assembly history of the galaxy with $M_\star = 10^{6.06}\,\mathrm{M_\odot}$ starts at $z \sim 7.5$ ($t^\mathrm{tot} \sim 500\,\mathrm{Myr}$), the first progenitor of the $M_\star = 10^{10}\,\mathrm{M_\odot}$ galaxy appears much earlier, at $z > 17$ ($t^\mathrm{tot} > 1\,\mathrm{Gyr}$). Comparing these lifetimes to the redshift of transition also shows that while the lowest mass galaxies spend a 100\% of their lifetime in the stochastic phase, the highest mass systems spend a negligible amount of time in the stochastic phase only at the highest redshifts (not shown in the plot). 

This plot already hints at two of the key results of the model: (i) the more massive a galaxy, the earlier it transitions from stochastic to steady star formation; and (ii) the more massive a galaxy, the larger is the fraction of its lifetime that it spends in the steady star forming phase.

\subsection{Characterizing the SFH} \label{characterizing}

We fit the SFH of the galaxies in the \textsc{astraeus} simulations with a simple redshift-dependent parametric form such that 
\begin{eqnarray}
    \label{eq:fit}
    \gamma(z, \alpha, \beta) = \mathrm{log}_{10}\left(\frac{\mathrm{SFR}(z)}{\msun/\mathrm{yr}}\right) &=& -\alpha \cdot (1 + z) + \beta,
\end{eqnarray}
where $\alpha$ determines the redshift-dependence of the SFH and $\beta$ is the normalization factor. This fit is performed using a least-square linear regression (orange solid line in Fig.~\ref{fig:6gal}).

In order to study halos with a reliable SFH, we only consider halos fulfilling the following criteria: (i) a minimum mass of a  $10^{8.95}\,h^{-1}\,\mathrm{M_\odot}$, corresponding to halos with at least a $100$ particles in our N-body simulation;  (ii) since performing a fit using Eqn.~\ref{eq:fit} is meaningful only if there is a minimum number of points in its SFH, we remove all galaxies that undergo star formation in less than 10 contiguous snapshots ($N\mathrm{_{SF}}$); this cut mainly removes low-mass galaxies that display a highly stochastic SFH at any given redshift. The impact of this choice is discussed in more detail in Appendix \ref{app:cuts}. After performing these selection cuts, the number of galaxies considered in this work are $\sim 1.6\times 10^6$ at $z=10$ and increase to $\sim 21\times10^6$ at $z=5$.

Next, we define a criterion for stochasticity: at any redshift, we assume that a galaxy is in the stochastic phase if its SFR deviates from Eqn.~\ref{eq:fit} by more than $\Delta_\mathrm{SFR}$. Throughout its life, a galaxy can alternate between periods of steady and stochastic star formation. We consider a galaxy to be in the stochastic star formation phase between the redshift of its formation and the lowest redshift at which it transitions from stochastic to non-stochastic star formation. We need to choose $\Delta_\mathrm{SFR}$ so that it is a sensible representation of the stochasticity - too low a value of $\Delta_\mathrm{SFR}$ leads to high-mass galaxies with steady star formation being categorized as stochastic. Alternatively, a value of $\Delta_\mathrm{SFR}$ that is too high leads to all galaxies being considered as steady star-formers. We choose $\Delta_\mathrm{SFR} = 0.6\,\mathrm{dex}$ as a reasonable compromise, indicated by the shaded area in Fig.~\ref{fig:6gal}. We discuss the impact of this choice in more detail in Appendix ~\ref{app:stoc}. Finally, we briefly discuss the dependence of the stochasticity criteria on the time and mass resolution of the underlying N-body simulation at the end of Sec.~\ref{trans} and Sec.~\ref{fitting}.

We note that this theoretical definition of stochastic star formation differs from estimates from observations \citep[e.g.][]{Faisst2019}. While we can follow the SFHs of galaxies at each epoch, observational estimates of SFHs are typically based on two data points (SFR in the last 10 and 100\,Myrs through measuring the H$\alpha$ line and UV luminosity, respectively) and a chosen SFH shape. The definition of stochastic star formation we use also differs from the definition introduced by \cite{caplar2019} who model the SFR of a galaxy as a stochastic process relative to the star-forming main sequence. Thus, while we define stochasticity in the SFH of a galaxy as a deviation from its own entire history, they define it as a deviation from the main sequence, which naturally occurs more often. Hence, we expect the definition we use to identify a lower fraction of galaxies as stochastic star forming galaxies, especially for low-mass galaxies ($M_\star < 10^9\msun$) where the variability of SFHs among galaxies is high (see Sec.~\ref{fitting}).

\begin{figure*}
    \centering
    \includegraphics[width=\textwidth]{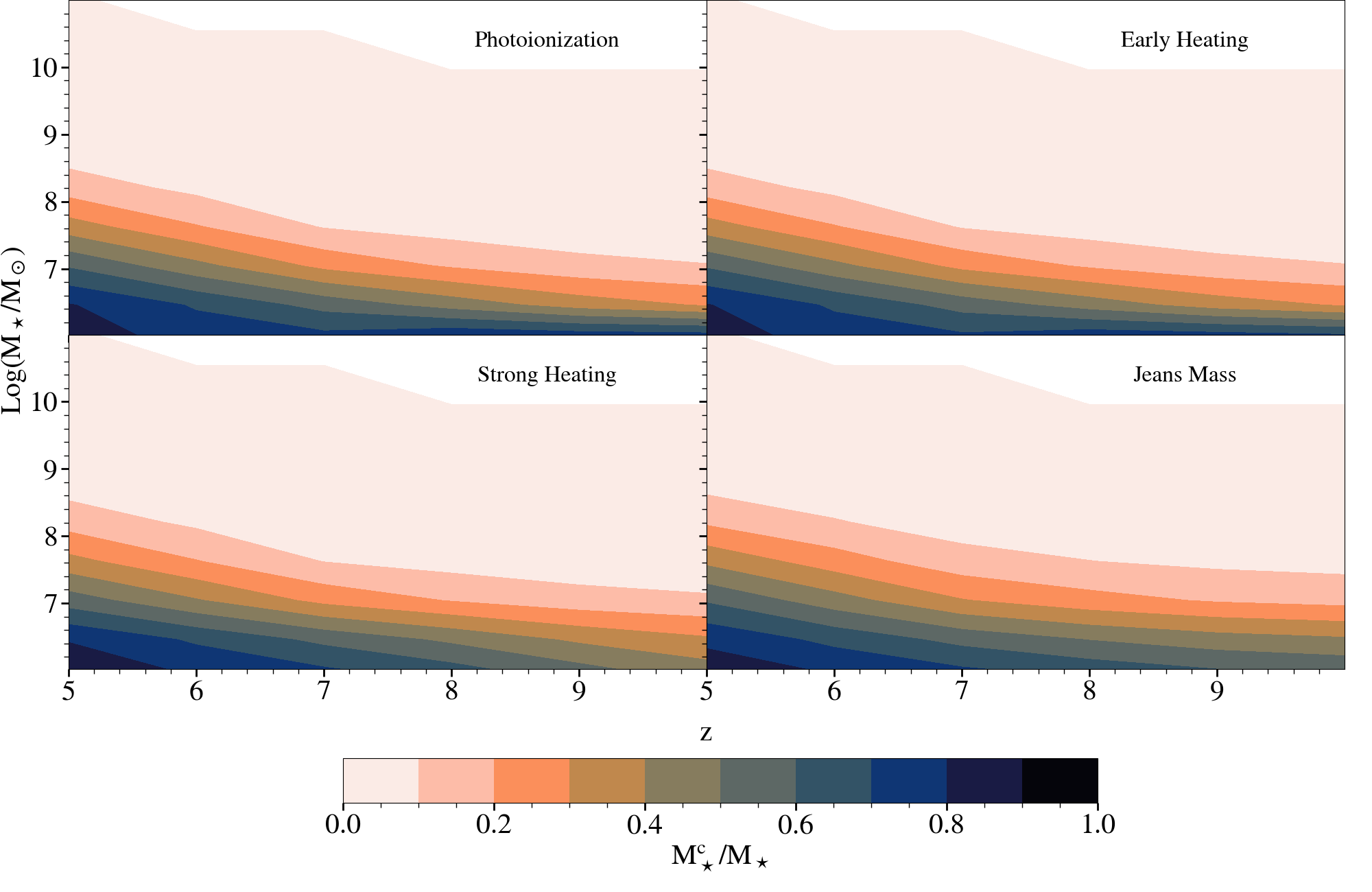}
    \caption{Mean fraction of stellar mass formed in the stochastic phase (color-coded as per the color-bar at the bottom) as a function of redshift and stellar mass for the 4 different radiative feedback models used in this work, as marked. }
    \label{fig:sto}
\end{figure*}

\begin{figure*}
    \centering
    \includegraphics[width=\textwidth]{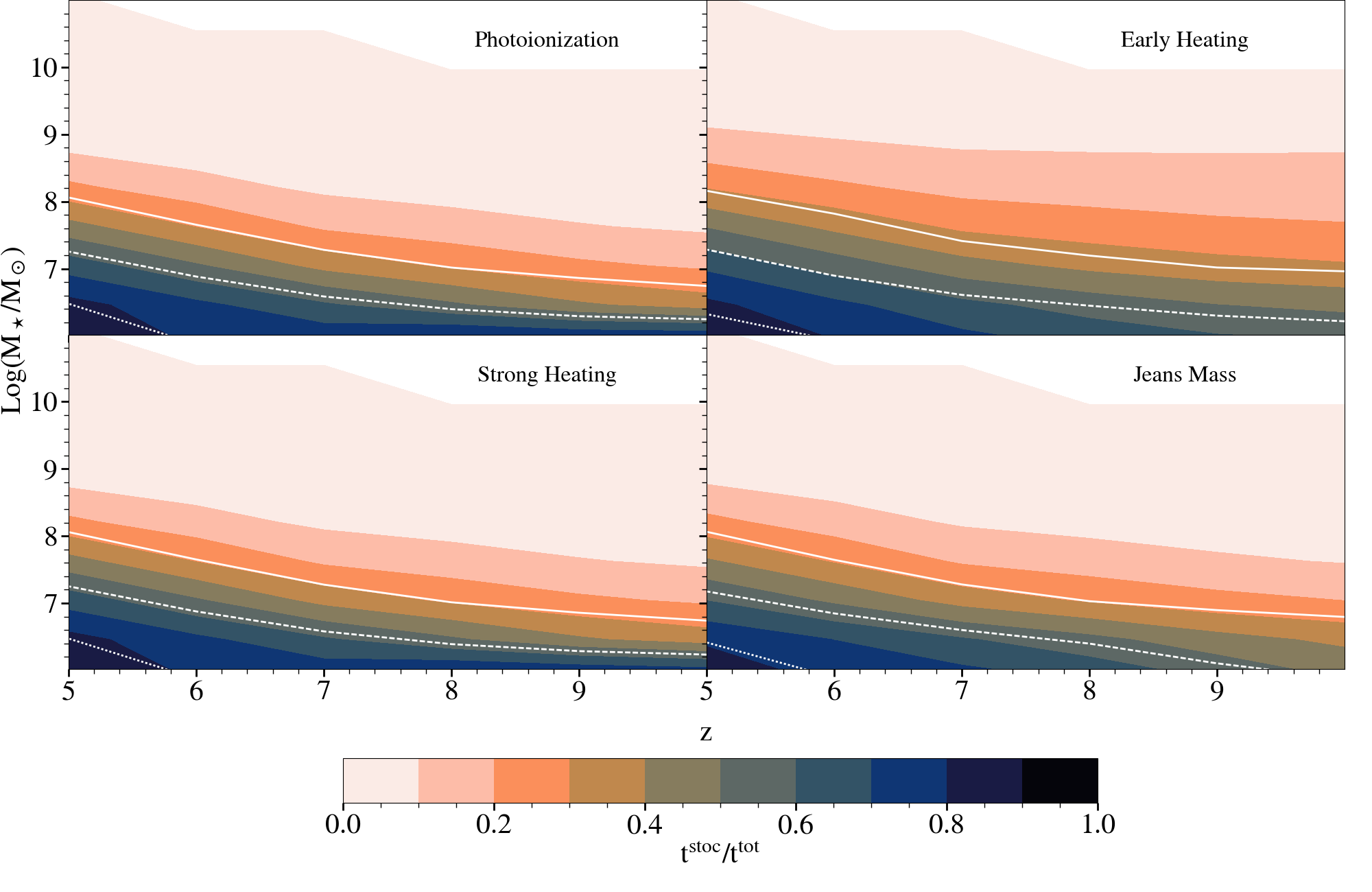}
    \caption{Mean fraction of time spent in the stochastic phase (color-coded as per the color-bar at the bottom) as a function of redshift and stellar mass for the 4 different radiative feedback models used in this work, as marked.  The solid, dashed and dotted lines represent the stellar mass below which 20\%, 50\% and 80\% of the stellar mass is formed in the stochastic phase, respectively. }
    \label{fig:timestoc}
\end{figure*}

\section{Quantifying the star formation histories} \label{Results}

In this section, we start by discussing the fractional stellar mass assembled and fractional lifetime spent in the stochastic phase in Sec. \ref{properties} before discussing the stellar mass and redshift of transition from stochastic to non-stochastic star formation in Sec. \ref{trans}. We end by showing the dependence of the fits to the SFHs on the stellar mass, redshift and radiative feedback model considered in Sec. \ref{fitting}.

\subsection{Fractional stellar mass assembled and lifetime spent in stochastic phase} 
\label{properties}

In this section, we start by discussing the mean fraction of stellar mass formed in the stochastic phase, which is expressed as $M_\star^\mathrm{c}/M_\star$ where $M_\star^\mathrm{c}$ represents the stellar mass assembled in the stochastic phase. The results of this calculation are shown in Fig.~\ref{fig:sto}. 

First, we consider the \textit{Photoionization} model shown in the top left panel. We can see that, at every redshift, a high fraction ($>70\%$) of the stellar mass contained in galaxies with $M_\star \simeq 10^{6.5}\,\mathrm{M_\odot}$ has been formed stochastically. These galaxies assemble in low-mass halos ($M_\mathrm{h} \sim 10^{9.5}\,\mathrm{M_\odot}$) that are \textit{feedback limited} (and hence highly stochastic in terms of star formation) throughout their history, since a small number of SN is enough to expel all of the gas in such systems. As the halo mass of a galaxy increases above the transition mass of $M_\mathrm{h} \sim 10^{9-9.3} \, \msun$ at $z \sim 5-10$ as discussed in Sec.~\ref{processes}, galaxies inherit increasingly more gas from their progenitors, most of which is kept bound within the halo and forms stars with a constant efficiency of $f_\star$. Hence, the fraction of stellar mass they form stochastically reduces with increasing halo (and stellar) mass. Indeed, by $z=5$, massive galaxies with $M_\star \gsim 10^{8.4}\,\mathrm{M_\odot}$ have formed at most 10\% of their stellar mass stochastically. Secondly, for a given value of $M_\star$, galaxies form a larger fraction of their stellar mass in the stochastic phase with decreasing redshift as seen from the same panel. For example, galaxies with $M_* \sim 10^{7.5}\msun$ form $<10\%\, (\sim 40\%)$ of their stellar mass stochastically at $z=10\, (5)$. This redshift trend can be explained as follows: due to their shallower potentials, galaxies of a given halo mass show lower effective star formation efficiencies at lower redshifts (as detailed in Sec. \ref{processes}). This results in galaxies of a given stellar mass residing in higher halo masses with decreasing redshift. The longer assembly histories (where low-mass progenitors are feedback limited), results in a larger fraction of the stellar mass forming stochastically. 

The same mass and redshift trends are seen for the three other reionization feedback models, as shown in the same figure. Just as the \textit{Photoionization} model, the \textit{Early Heating} model has a time-delayed, weak radiative feedback and thus yields very similar results, as we can see in the top right panel of Fig.~\ref{fig:sto}. For the \textit{Strong Heating} model, the time-delayed nature of the feedback leads to similar results as for the \textit{Photoionization} model above $M_\star > 10^{6.4}\,\msun$. However, at lower stellar masses above $z=7$, the fraction of stellar mass formed stochastically is slightly lower in the \textit{Strong Heating} model. Although we would expect the stochasticity to increase in the presence of stronger radiative feedback, the choice of $N_{SF}=10$ introduces a bias by removing galaxies that are too stochastic, hence lowering the average fraction of stellar mass formed stochastically. The \textit{Jeans Mass} model is shown in the bottom right panel and we can see that the fraction of stellar mass formed stochastically increases by up to 10\% in galaxies between $M_\star \sim 10^7\,\mathrm{M_\odot}$ and $M_\star \sim 10^{8.6}\,(10^{7.8})\,\mathrm{M_\odot}$ at $z=5\,(10)$. The instantaneous nature of the radiative feedback leads to a strong gas suppression in their progenitors, which coupled with SN feedback, increases the stochasticity. Additionally, for the same reason as the \textit{Strong Heating} model, galaxies with $M_\star \leq 10^{6.4}\,\mathrm{M_\odot}$ above $z=7$ form a lower fraction of their stellar mass in the stochastic phase compared to the \textit{Photoionization} model.

The trends seen in Fig.~\ref{fig:sto} are reflected in Fig.~\ref{fig:timestoc} which shows the mean fraction of lifetime spent in the stochastic phase $t^\mathrm{stoc}/t^\mathrm{tot}$. Here, $t^\mathrm{stoc}$ and $t^\mathrm{tot}$ are the time spent in the stochastic phase and the total lifetime (as defined is Sec.~\ref{starform}), respectively. Firstly, for each model, the fraction of time spent in the stochastic phase scales with the fraction of stellar mass formed in that phase and thus is higher for galaxies with a shallower gravitational potential. For example, it decreases with stellar mass from $90\%$ at $M_\star \simeq 10^{6}\,\mathrm{M_\odot}$ to less than $10\%$ at $M_\star> 10^{9.2}\,\mathrm{M_\odot}$ at $z=5$. Secondly, for a given stellar mass, $t^\mathrm{stoc}/t^\mathrm{tot}$ increases with decreasing redshift. As noted above, galaxies of a given stellar mass are hosted in progressively more massive halos with decreasing redshift. Their longer assembly histories from feedback-limited low-mass progenitors leads to an increase in the fractional lifetime spent in the stochastic star forming phase. For example, in all the models shown here, galaxies of $M_\star \simeq 10^{7}\,\mathrm{M_\odot}$ show an increasing $t^\mathrm{stoc}/t^\mathrm{tot}$ value from $\lsim 30\%$ at $z \sim 10$ to $\sim 50\%$ by $z \sim 5$. Comparing the different radiative feedback models, the first two models are here again very similar. As expected, the lower stochasticity for galaxies with $M_\star \leq 10^{6.4}\,\mathrm{M_\odot}$ above $z=7$ in the \textit{Strong Heating} and \textit{Jeans Mass} models is correlated with a lower fraction of time spent in the stochastic phase. In addition for the \textit{Jeans Mass} model, $t^\mathrm{stoc}/t^\mathrm{tot}$ in galaxies between $M_\star \sim 10^7\,\mathrm{M_\odot}$ and $M_\star \sim 10^9\,\mathrm{M_\odot}$ is higher by up to 10\% compared to all other models. It follows the same trend as the fraction of stochastic stellar mass formed from $M_\star \sim 10^7\,\mathrm{M_\odot}$ to $M_\star \sim 10^{8.6}\,(10^{7.8})\,\mathrm{M_\odot}$ at $z=5\,(10)$. However, between $M_\star \sim 10^{8.6}\,(10^{7.8})\,\mathrm{M_\odot}$ at $z=5\,(10)$ and $M_\star \sim 10^9\,\mathrm{M_\odot}$, the increase in $t^\mathrm{stoc}/t^\mathrm{tot}$ does not reflect an increase in $M_\star^\mathrm{c}/M_\star$. Here, the stronger radiative feedback causes a stronger suppression of star formation, resulting in these galaxies remaining longer in the stochastic phase while building up stellar mass at a lower rate than in all other radiative feedback models.

The trend of stochastic star formation being prevalent in lower mass galaxies is in rough agreement with observational findings. Using a sample of observed galaxies with stellar masses of $10^{8.5}\msun< M_\star < 10^{11.5}\msun$ at $z\sim4.5$, \cite{Faisst2019} find that the excess of H$\alpha$ luminosity compared to UV luminosity decreases with increasing stellar mass, indicating a reduced burstiness of the SFH of massive galaxies. 
However, in their sample, high mass galaxies ($M_\star > 10^{10}\msun$) show signs of recent bursts of star formation, which are not present in our simulations. A possible explanation for this difference comes from the fact that \citet{Faisst2019} compare SFRs within the last $\sim10\,$Myr to SFRs within the last $\sim100\,$Myr, while the time steps of our simulation exceed $10\,$Myr at $z\simeq5$ ($\Delta t \simeq35\,$Myr), resulting in any burst lasting less than $35\,$Myr being smoothed out. 
Similarly, measuring the timescale on which the SFR in a galaxy loses "memory" of previous star formation, \cite{caplar2019} suggest that the stochasticity of the SFH decreases with increasing stellar mass for $z=0$ galaxies with $M_\star < 10^{10}\,\msun$.

\subsection{The redshift and stellar mass of transition to steady star formation}
\label{trans}
\begin{figure*}
    \centering
    \includegraphics[width=\textwidth]{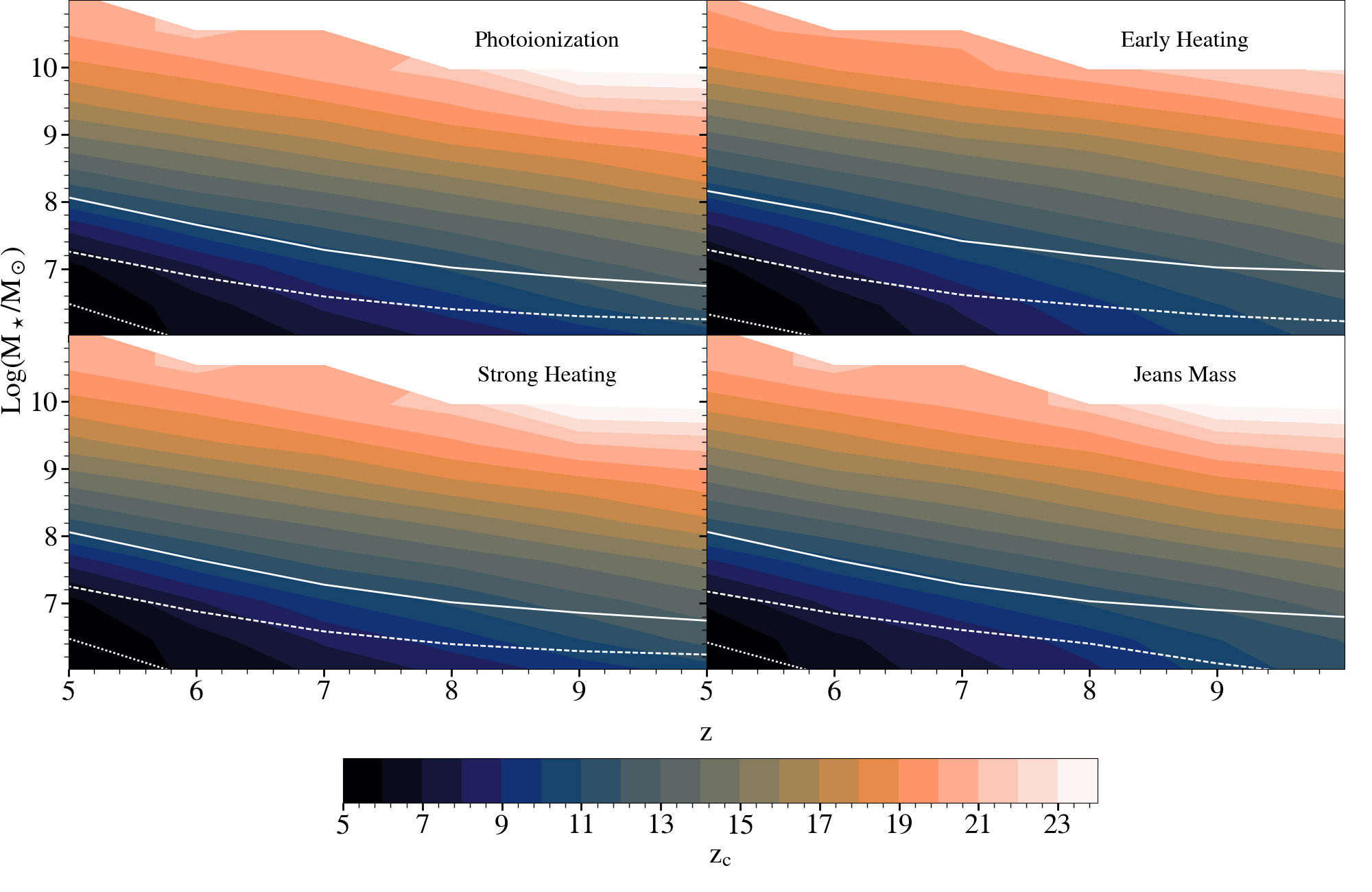}
    \caption{Median critical redshift $z_\mathrm{c}$ (color-coded as per the color-bar at the bottom) as a function of redshift and stellar mass for the 4 different radiative feedback models used. The solid, dashed and dotted lines represent the stellar mass below which 20\%, 50\% and 80\% of the stellar mass has been formed in the stochastic phase, respectively.}
    \label{fig:zcrit}
\end{figure*}

\begin{figure*}
    \centering
    \includegraphics[width=\textwidth]{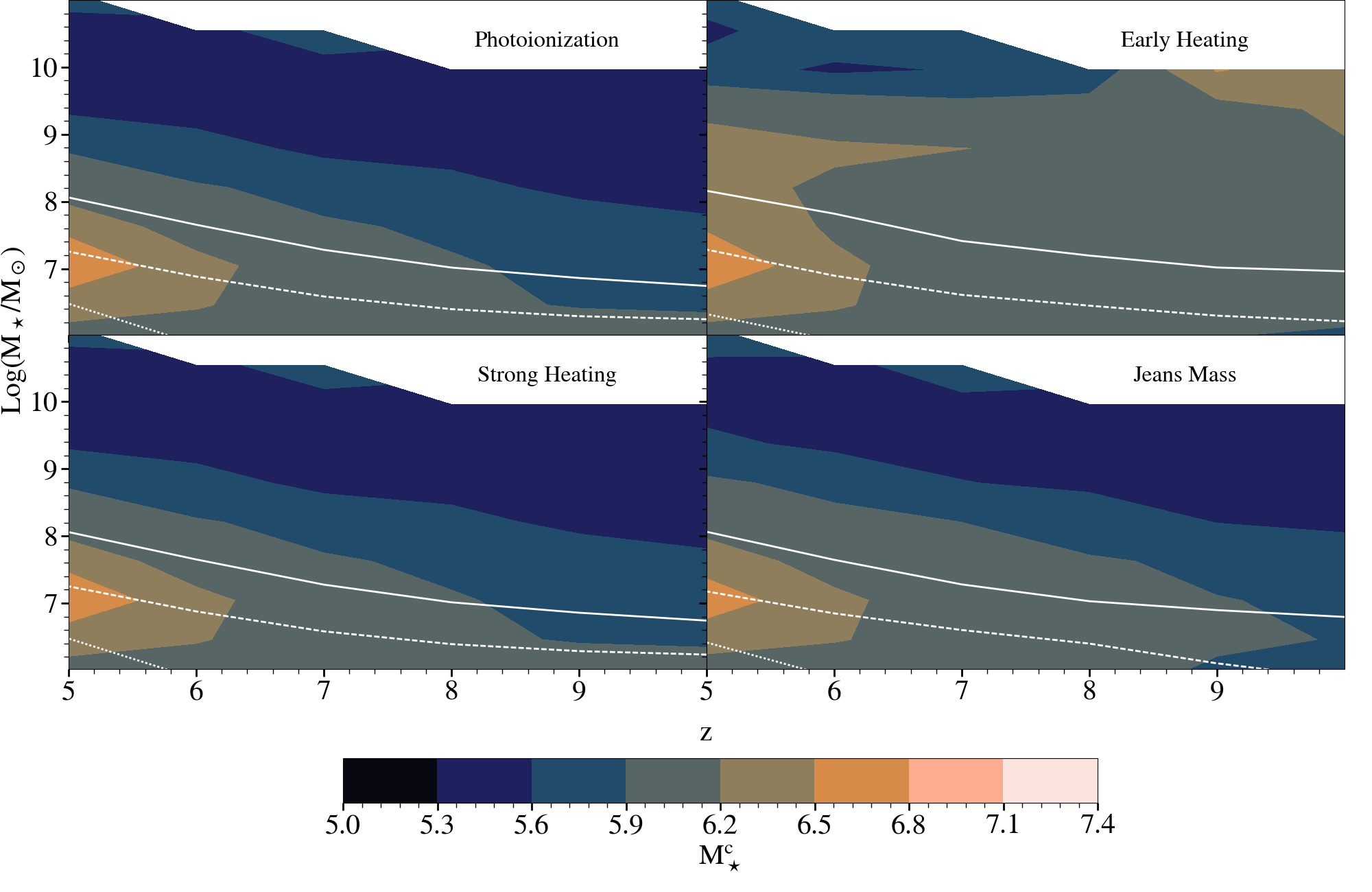}
    \caption{Median critical mass (color-coded as per the color-bar at the bottom) as a function of redshift and stellar mass for the 4 different radiative feedback models used. The solid, dashed and dotted lines represent the stellar mass below which 20\%, 50\% and 80\% of the stellar mass has been formed in the stochastic phase, respectively.}
    \label{fig:mcrit}
\end{figure*}

To reliably reconstruct the star formation assembly of galaxies, we need to assess the redshift below which our fit (Eqn. \ref{eq:fit}) is an accurate representation of the underlying SFH. We show the median of this critical redshift ($z_\mathrm{c}$; Fig.~\ref{fig:zcrit}) and the associated median critical stellar mass ($M_\mathrm{\star}^\mathrm{c}$; Fig.~\ref{fig:mcrit}) at which galaxies transition from the stochastic to steady star forming phase as a function of stellar mass at $z=5-10$. Starting with Fig.~\ref{fig:zcrit} we see that, at every redshift, the duration of steady star formation increases with the stellar mass. For example, in the \textit{Photoionization} model shown in the first panel, galaxies with $M_\star = 10^{10}\,\mathrm{M_\odot}$ at $z=5$ escape the stochastic phase at $z_c \sim 19$, while galaxies with $M_\star = 10^{6}\,\mathrm{M_\odot}$ become non-stochastic only at $z_c \sim 6$. This is partly due to their longer lifetime and partly due to their deeper potentials that allow them to have a steady SFH (see bottom panels of Fig.~\ref{fig:6gal}). As they escape the stochastic phase at an earlier cosmic time (i.e at a deeper potential well for a given stellar mass), they build a lower mass in this phase, which leads to $M_\mathrm{\star}^\mathrm{c}$ decreasing with increasing stellar mass. 

We would expect that $M\mathrm{_\star^c}$, shown in Fig. \ref{fig:mcrit}, could be derived by multiplying the average fraction of stellar mass formed in the stochastic phase, $\langle M_\star^\mathrm{c}/M_\star \rangle$, with the stellar mass. However, we find $M_\star^\mathrm{c}$ to be lower than what we would derive from such a calculation. For example, for galaxies with $M_\star \sim 10^{8.4}\msun$, the value of $M_\star^\mathrm{c}$ in Fig. \ref{fig:sto} suggests a value of $\langle M_\star^\mathrm{c} \rangle \sim 10^{7.4}\msun$, while we find a value of $M_\star^\mathrm{c} \sim 10^{6}\msun$ in Fig. \ref{fig:mcrit}. The reason for this deviation is the positively skewed distribution of $M\mathrm{_\star^c}$, which results in the median (shown in Fig. \ref{fig:mcrit}) being lower than the mean (shown in Fig. \ref{fig:sto}). A few galaxies with a longer stochastic star formation phase than most other galaxies in the chosen stellar mass bin can increase the mean $M\mathrm{_\star^c}$ value considerably, while the median $M\mathrm{_\star^c}$ value remains unaffected. Moreover, we note that we only consider the combined SFH of all progenitors of a galaxy. Hence, summing up the stochastic star formation events in individual progenitors can lead to an apparent non-stochastic star formation in the combined SFH, leading to a lower $M\mathrm{_\star^c}$ value. In this respect, the $M\mathrm{_\star^c}$ values represent a lower limit of the stellar masses formed stochastically.

We note that, unlike $\langle M_\star^\mathrm{c}/M_\star \rangle$ which decreases with stellar mass at all redshift, the median value of $M_\star^\mathrm{c}$ increases with stellar mass for galaxies below $M_\star = 10^7\,\mathrm{M_\odot}$ at $z\simeq5-6$. This is a consequence of a high fraction of galaxies still being in the stochastic phase as shown by the white dashed line representing the limit below which more than 50\% of the stellar mass has been formed stochastically. Hence, the critical mass is skewed towards the current stellar mass and increases with it. We see that the first three models give similar results except for the slight decrease of the critical mass in \textit{Strong Heating} model for galaxies below $10^{6.4}\,\mathrm{M_\odot}$ at $z=9$ and $z=10$. However, in the \textit{Jeans Mass} model, we see that galaxies above $10^8\,\mathrm{M_\odot}$ escape the stochastic phase later and with a higher mass, e.g. galaxies with $M_\mathrm{h} \sim 10^{10}\,\mathrm{M_\odot}$ at $z=10$ going from $M_\star^\mathrm{c} \sim 10^{5.6}\,\mathrm{M_\odot}$ ($z_c \sim 23$) in the \textit{Photoionization} model to $M_\star^\mathrm{c} \sim 10^{6.2}\,\mathrm{M_\odot}$ ($z_c \sim 21$) in the \textit{Jeans Mass model}. Due to its instantaneous nature, the radiative feedback in this model strongly impacts these high-mass galaxies early in their history when they had a shallower potential well and ends up delaying their transition to steady star formation to a later time/higher stellar mass.

To summarise, the emerging picture of the evolution of the SFHs of galaxies above $z=5$ is as follows: at all redshifts, low-mass galaxies form most of their stellar mass stochastically. As they become more massive through mergers and accretion, their gravitational potential deepens and they can convert a higher fraction of their gas into stars, becoming steady star formers. Further, as galaxies of a given stellar mass have increasingly shallower potentials with decreasing redshift, the average transitional stellar mass between the stochastic and steady star forming phases increases with decreasing redshift as does the time required to build that mass. The choice of radiative feedback has a limited impact on the stochasticity: only the strongest model (\textit{Jeans Mass}) increases the stellar mass and time needed to transit out of the stochastic star formation phase significantly.

To assess the dependence of the stochasticity criteria on the time and mass resolution of the underlying N-body simulation, we carry out a resolution test using the \textit{Extremely Small MultiDark Planck} (\textsc{esmdpl}) simulation. The \textsc{esmdpl} simulation has a smaller box size of $64\, h^{-1}\mathrm{cMpc}$ and a 20$\times$ higher mass resolution (DM particle mass of $3.3\times10^5\,h^{-1}\msun$) than the \textsc{vsmdpl} simulation. As we increase the mass resolution, lower mass galaxies are resolved at all redshifts and the emergence of their first progenitors shifts to earlier times, leading to longer lifetimes. Hence, compared to the \textsc{vsmdpl} simulation, we find the fractional lifetimes spent and stellar mass assembled in the stochastic phase for galaxies with $M_\star < 10^9\, \msun$ to increase by up to $\sim20\%$ and $\sim10\%$ in the \textsc{esmdpl} simulation, respectively. These two quantities are similar for galaxies with $M_\star > 10^9\, \msun$ in both simulations. Nevertheless, when we consider galaxies for which the halo mass functions of the \textsc{esmdpl} and \textsc{vsmdpl} simulations converge ($M_\mathrm{h}\simeq10^{8.6}-10^{10}\msun$ at $z=9$ and $M_\mathrm{h}\simeq10^{8.6}-10^{11}\msun$ at $z=6$), we find the $M_\star^\mathrm{c}$ values of galaxies in the \textsc{esmdpl} simulation to be in rough agreement with those obtained from the \textsc{vsmdpl} simulation.

\subsection{Fitting the SFH} \label{fitting}

Using the method described in Sec.~\ref{characterizing}, we then fit the SFH of each galaxy following Eqn.~\ref{eq:fit} and recover its slope, $\alpha(M_\star, z)$, and normalization, $\beta(M_\star, z)$. We start by showing $\alpha(M_\star, z)$ at $z=5-10$, for the four different radiative feedback models studied, in Fig.~\ref{fig:alpha}. The solid lines represent the fits at different redshifts, presented in Appendix~\ref{Appendix1}. We note, since the SFHs of low-mass galaxies are too stochastic to be accurately fitted by a single parametric law, these fits account only for galaxies for which at least 80\% of the stellar mass has been assembled in the steady star formation phase; this mass, $M_\mathrm{stoc}$, is represented by the vertical lines in the same plot. We see that in the \textit{Photoionization, Early and Strong} models, the slope of the SFH of low-mass galaxies, with $M_\star\leq 10^8\,(10^9)\mathrm{M_\odot}$ at $z=10\,(5)$, increases with stellar mass, due to the rapid increase of $M_\mathrm{g}^\mathrm{i}$ (Fig.~\ref{fig:feffMgasIni}). As we go to higher stellar masses of $M_\mathrm{h} > 10^8\, (10^{9})\,\mathrm{M_\odot}$ at $z=10\,(5)$ and both the cumulative SN and radiative feedback remove only a negligible fraction of the gas mass. In this case, $M_\mathrm{g}^\mathrm{i}$ essentially scales with the halo mass, that steadily assembles as $\propto -(0.2\sim0.25)z$, leading to a constant slope of $\sim 0.18$\footnote{We note that the SFHs show shallower slopes than the dark matter mass assembly histories due to the star formation stochasticity in the early SFHs of galaxies. On average, the latter slightly reduces the slope of the SFH and is not present in the dark matter assembly.}
The reduction of the scatter for $\alpha(M_\star, z)$ as $M_\star$ increases shows that, although low-mass galaxies display a variety of SFH, their assembly histories converge as their masses increase. 

While the \textit{Jeans Mass} model shows the same slope and reduced scatter for high mass galaxies as the other models, $\alpha$ decreases with increasing stellar mass just above $M_\mathrm{stoc}$ at $z \geq 8$. This result is to be weighed by the fact that there is a big dispersion in the SFH slope at the low-mass end. Unlike in the other models, in the \textit{Jeans Mass} model the gas mass in a galaxy is immediately reduced as soon as the surrounding region is ionized. This instantaneous feedback leads to a strong reduction of star formation especially in the early history of galaxies. Hence, the slope of the SFH increases compared to the other models.

\begin{figure*}
    \centering
    \includegraphics[width=\textwidth]{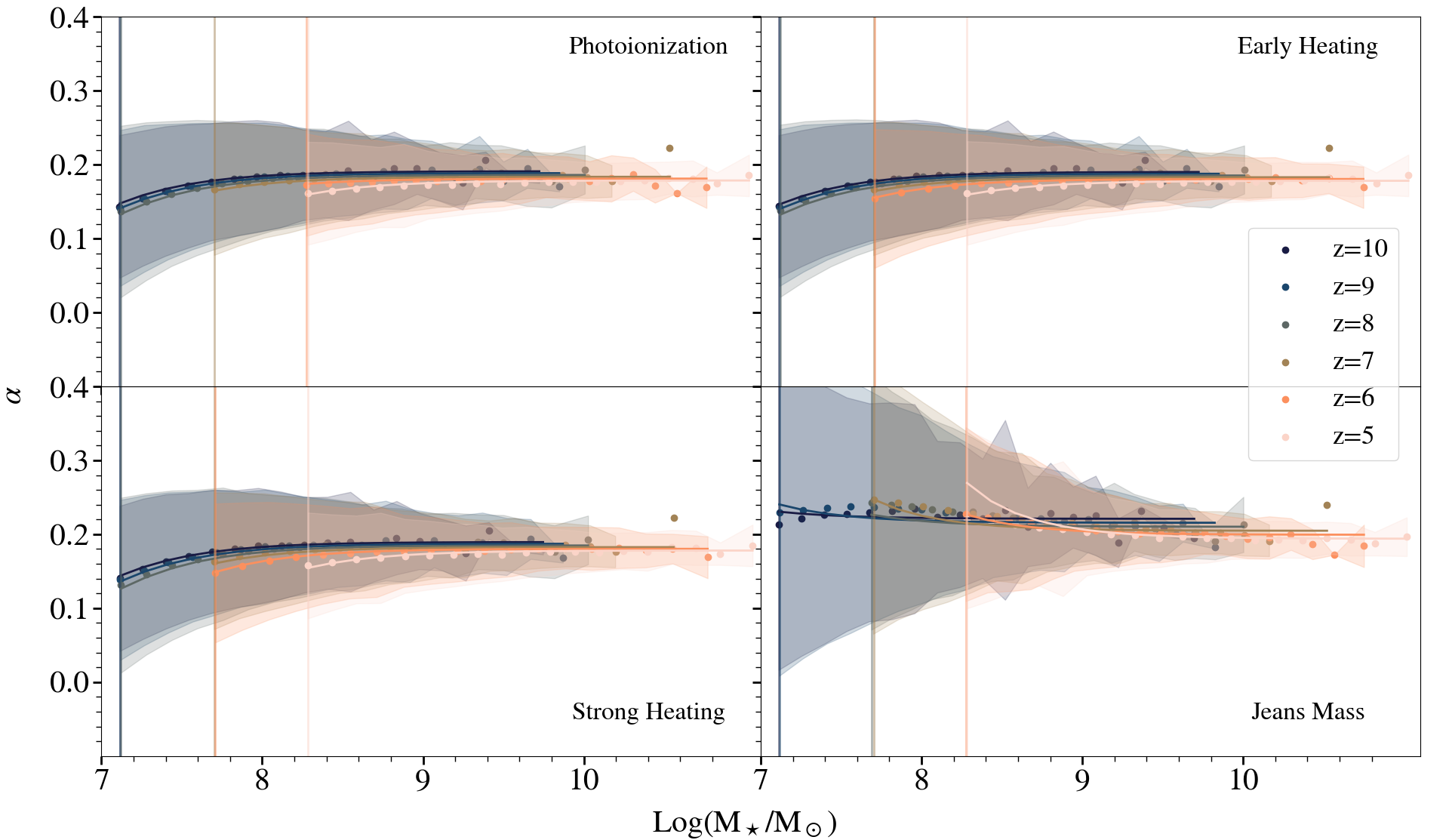}
    \caption{Mean slope ($\alpha$) of the SFH of galaxies as a function of stellar mass at $z=5-10$ (as marked) for the 4 different radiative feedback models used in this work. The vertical lines demarcate the stellar mass above (below) which less (more) than 20\% of the stellar mass has been formed stochastically. The dots represent the results of the simulation while the lines are the fits. The fits for $\alpha$ are shown in Appendix \ref{Appendix1}.}.
    \label{fig:alpha}
\end{figure*}

\begin{figure*}
    \centering
    \includegraphics[width=\textwidth]{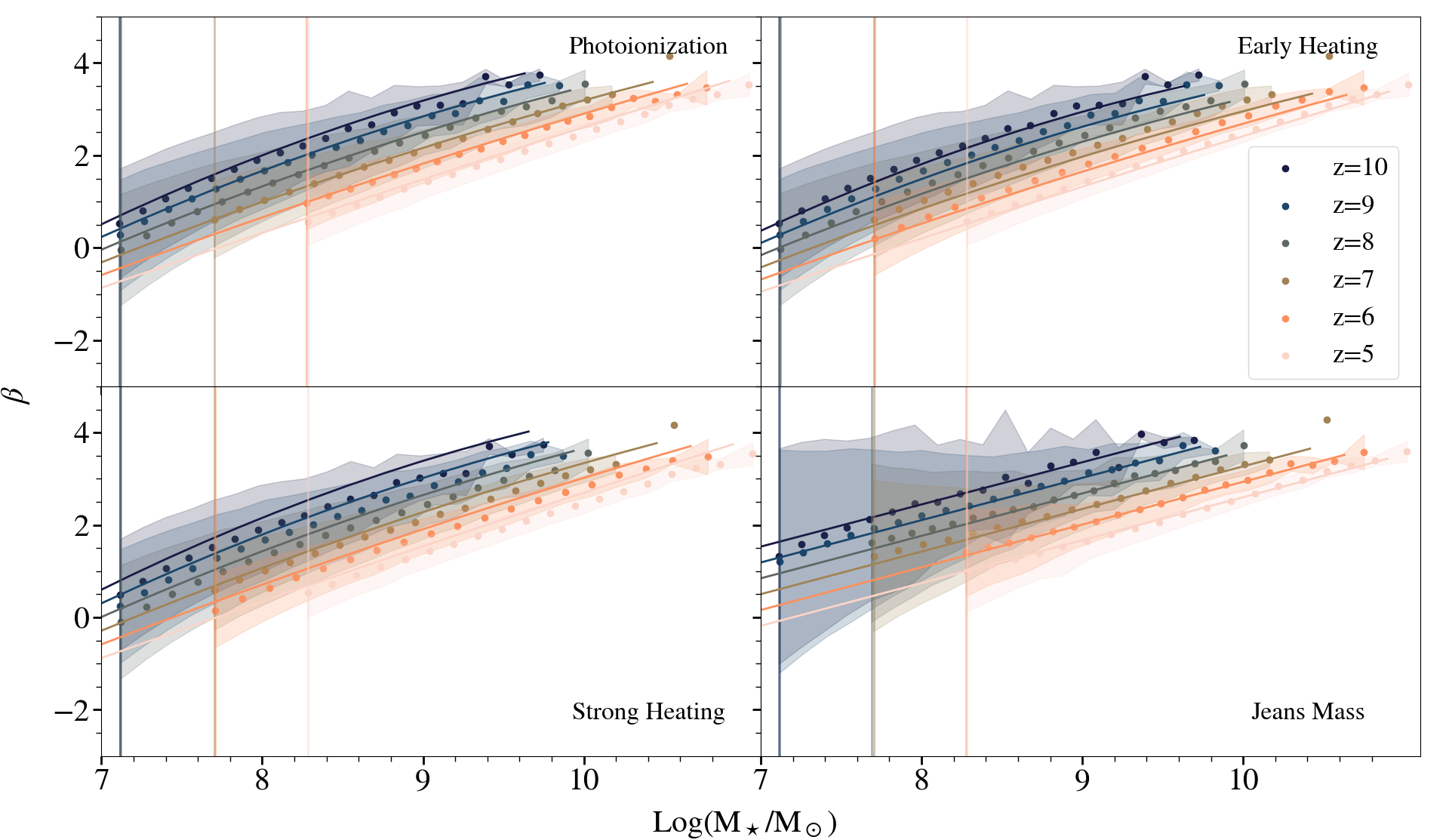}
    \caption{Mean normalisation ($\beta$) of the SFH of galaxies as a function of stellar mass at $z=5-10$ (as marked) for the 4 different radiative feedback models used in this work. The vertical lines demarcate the stellar mass above (below) which less (more) than 20\% of the stellar mass has been formed stochastically. The dots represent the results of the simulation while the lines are the fits. The fits for $\beta$ are shown in Appendix \ref{Appendix1}.}
    \label{fig:beta}
\end{figure*}

\begin{figure}
    \centering
    \includegraphics[width=\columnwidth]{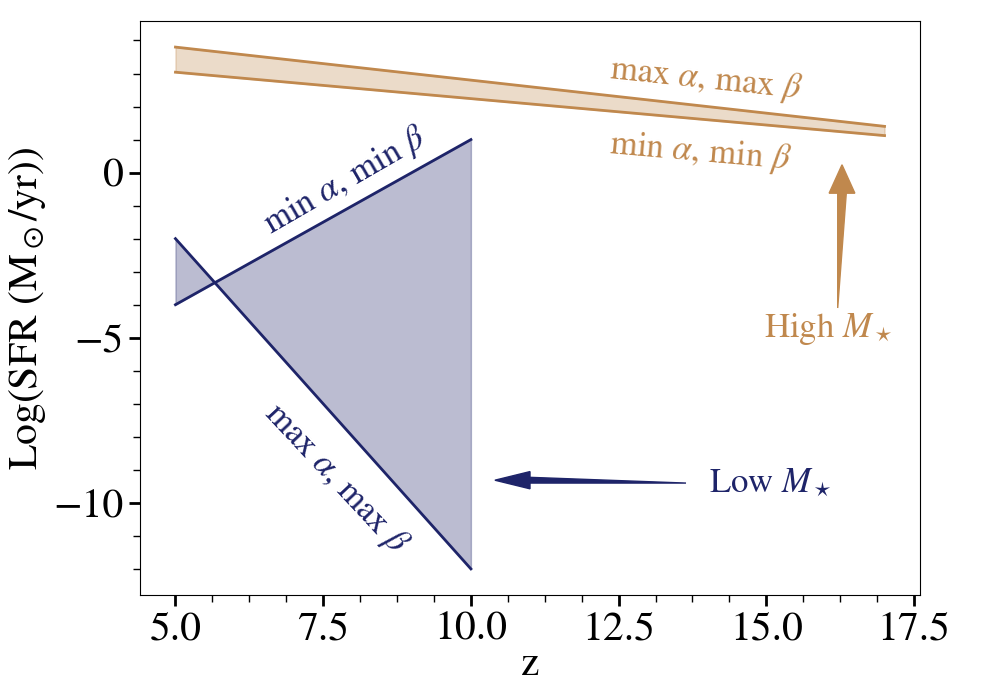}
    \caption{Schematic illustrating the range of possible SFHs for galaxies as a function of their stellar mass and redshift. While high-mass galaxies have similar SFHs, low-mass galaxies display a wide range of SFHs, from an initial burst followed by a declining SFH to an early low SFR that rapidly increases with redshift.}
    \label{fig:sketchalphabeta}
\end{figure}

Considering the normalization of the SFH (Fig.~\ref{fig:beta}), at a given redshift, the mass-independent slope of high-mass galaxies, with $M_* \geq 10^8\,(10^9)\mathrm{M_\odot}$ at $z=10\,(5)$, naturally results in $\beta(M_\star, z)$ scaling positively with the stellar mass. 
For lower stellar masses, $\beta(M_\star, z)$ scales strongly with $\alpha(M_\star, z)$, leading to a variety of assembly histories (see blue region in Fig.~\ref{fig:sketchalphabeta}): a galaxy of mass $M_\star <  10^8\,(10^9)\mathrm{M_\odot}$ at $z=10\,(5)$ can build up its stellar mass either by (i) an early starburst that is followed by a declining star formation rate in the absence of gas accretion i.e. a negative $\alpha(M_\star, z)$ and low $\beta(M_\star, z)$; (ii) forming stars at a somewhat constant rate when gas heating/ejection through feedback processes and gas accretion balance each other i.e. null $\alpha(M_\star, z)$, intermediate $\beta(M_\star, z)$; or (iii) forming increasingly more stars over time as gas accretion dominates i.e. positive $\alpha(M_\star, z)$, high $\beta(M_\star, z)$. Towards lower masses, SN and radiative feedback become more efficient in preventing star formation, leading to a flattening of the average SFH and thus a decrease in $\alpha(M_\star, z)$ as well as to a decrease in $\beta(M_\star, z)$ due to $\alpha(M_\star, z)$ and $\beta(M_\star, z)$ being correlated. It is also this correlation and the ability of feedback to suppress star formation completely that explain the increase of scatter with decreasing stellar mass. The fit of $\beta$ is presented in Appendix~\ref{Appendix1}.

Additionally, at every stellar mass, $\beta(M_\star, z)$ increases with increasing redshift, e.g going from a value of $1$ at $z=5$ to $3$ at $z=10$ for galaxies with $M_\star = 10^9\,\mathrm{M_\odot}$. As noted in previous sections, this is due to the fact that a galaxy of a given stellar mass has a higher halo mass at higher redshifts, resulting in a higher value of the SFR throughout its history.

Fig.~\ref{fig:beta} also shows that the choice of radiative feedback has little effect on $\beta(M_\star, z)$ above $M_\mathrm{stoc}$. Here again, only the \textit{Jeans Mass} model shows a noticeable difference for galaxies with stellar masses close to $M_\mathrm{stoc}$. Furthermore, the correlation between $\alpha(M_\star, z)$ and $\beta(M_\star, z)$ combined with the increase of $\alpha(M_\star, z)$ with decreasing stellar mass at the low-mass end leads to a higher $\beta(M_\star, z)$ in this model.

We note that applying the same stellar mass cut as in the \textsc{vsdmpl} simulation to the \textsc{esmdpl} simulation ($M_\star > M_{\rm stoc}$) yields $M_{\rm stoc} \sim 10^{7.7}\,(10^{8.8})\msun$ at $z=10\,(5)$. In this mass range, we find that the $\alpha$ and $\beta$ values are slightly higher for galaxies at $z=6-10$ in the \textsc{esmdpl} simulation but remain within the $1\sigma$ uncertainties shown as the shaded area in Fig.~\ref{fig:alpha} and \ref{fig:beta}.

We also compare our results to those obtained with empirical models \citep[][]{Behroozi2013}, hydrodynamical simulations assuming a uniform UVB \citep[][]{Finlator_2010} and radiative hydrodynamical simulations \citep[][]{Ocvirk2020}.
Using an empirical model that populates the halos of an N-body simulation with galaxies and is constrained by the observed stellar mass functions, the specific star formation rates of galaxies and the cosmic star formation rate at $z=0-8$, \citet{Behroozi2013} find the best-fitting SFHs of galaxies with $M_\mathrm{h} \geq 10^{11}\msun$ at $z\geq3$ to scale as $\mathrm{SFR}(t) \propto t^{3-4}$. These SFHs are steeper than the SFHs derived from the \textsc{astraeus} simulations (resulting in galaxies of a given $M_\star$ having younger stellar populations and thus a higher UV magnitude), since the SFHs in \citet{Behroozi2013} consider only the main branch while our SFHs represent the sum of the SFHs of all progenitors of a galaxy. While the recent SFR of a massive galaxy ($M_\mathrm{h} \geq 10^{11}\msun$) is dominated by its main branch, the SFRs of its progenitors during the early phases of its assembly have similar values. Accounting only for the main branch results in a SFH with lower SFRs during the initial phase of the galaxy's mass assembly and leads to a steeper slope of the SFH. For this reason, we find works that include the star formation of all progenitors in a galaxy's SFH to be in better agreement with our results. For instance, from their hydrodynamical simulations \citet{Finlator_2010} find the average SFHs of galaxies with $M_\star \geq 10^{8.2}\msun$ at $z\geq 5$ to follow $\mathrm{SFR}(t) \propto t^{1.7}$, which is also echoed by the evolution of the cosmologically averaged SFRs derived from observations in \cite{Papovich2011}. Similarly, the average SFHs of $M_\mathrm{h}\simeq10^{10-11}\msun$ galaxies in the radiative hydrodynamical simulation \textsc{codaii} \citep{Ocvirk2020} exhibit slopes that are very similar to those found in our simulations. Fitting their published SFHs using our methodology yields $\alpha(M_\star, z) \sim 0.2$ for $M_\mathrm{h}\,(M_\star) \geq 10^{10}\,(10^{8.5})\msun$. These and our results are also in agreement with the SFHs found for lower redshift ($z<5$) galaxies, yielding increasing SFHs during the EoR \citep[e.g][]{Diemer_2017, Ciesla2017}. However, the increasing SFH slope of a $z<5$ galaxy at $z>6$ depends on its detailed mass assembly history: while a galaxy with a  major merger at $z<5$ will exhibit a shallower SFH slope at $z>5$, a galaxy with (a) minor merger(s) at $z<5$ will show a steeper SFH at $z>5$. For this reason, the exact shape of the $z>5$ SFHs including all progenitors depends on whether the SFHs of low- or high-redshift galaxies are considered.

\section{Validation of the model} \label{Validation}

We have quantified the SFHs of galaxies above $z=5$ and derived a general formula (Eqn.~\ref{eq:fit}) and parameters (Appendix~\ref{Appendix1}) to express them as a function of stellar mass and redshift in the different radiative feedback scenarios explored in this work. In this section, we assess the capacity of our fits to recover two key properties of galaxies: their stellar mass and UV magnitude. 

\begin{figure*}
    \centering
    \includegraphics[width=\textwidth]{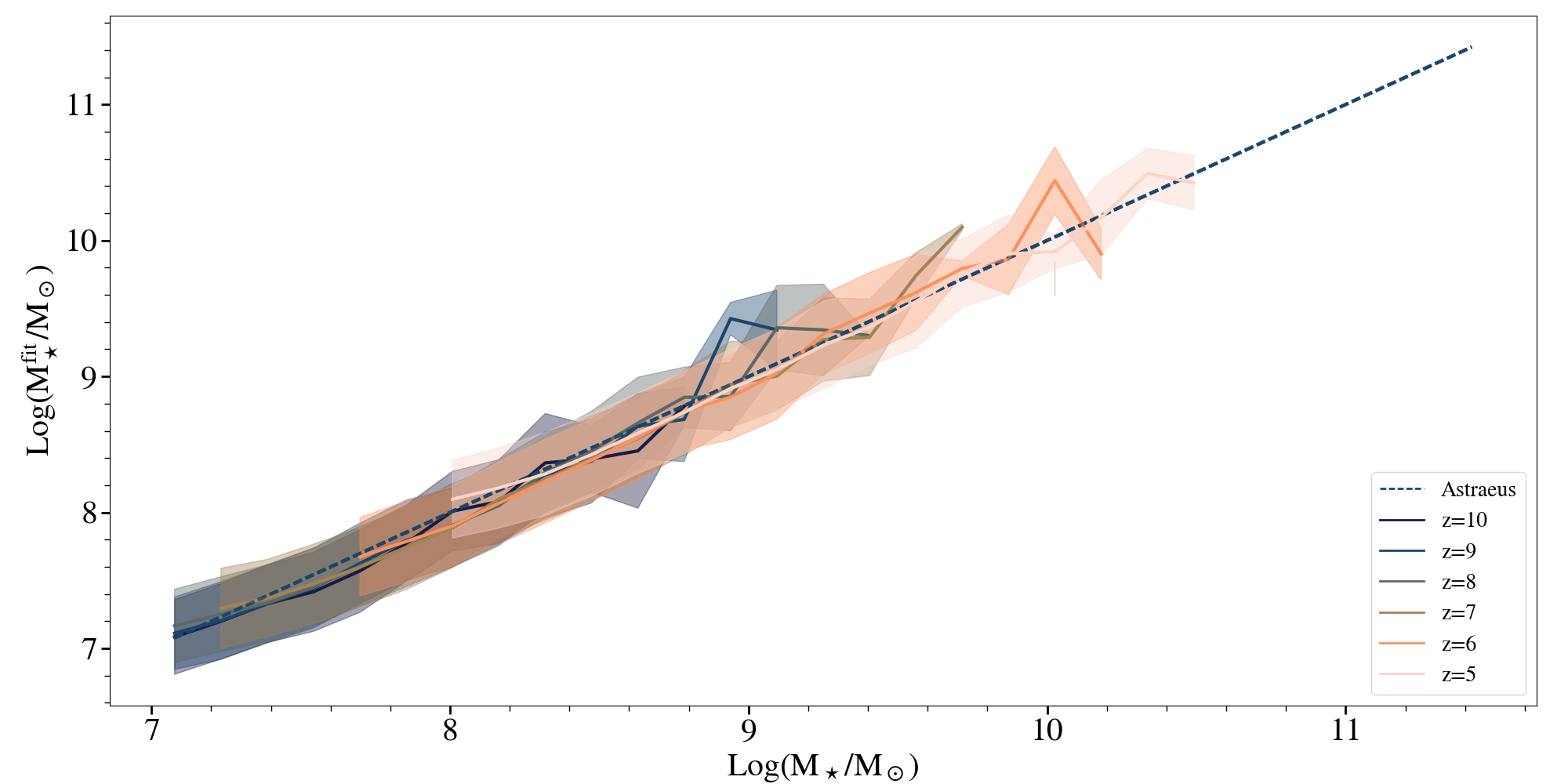}
    \caption{Median predicted stellar mass ($M_\star^\mathrm{fit}$) as a function of the stellar mass directly obtained from the \textsc{astraeus} simulation at $z=5-10$ for the \textit{Photoionization} model. The shaded area represents the standard deviation obtained by assuming a Gaussian spread ($\sigma = 0.3$) around the average value of $\beta(z, M_\star)$.}
    \label{fig:msfitted}
\end{figure*}

For each galaxy of mass $M_\star$ observed at redshift $z$, we compute the stellar mass predicted by integrating our fitted SFH ($M_\star^\mathrm{fit}$)
\begin{align}
    M_\star^\mathrm{fit}(M_\star,z) & = M^c_\star(M_\star,z) + \int_{z_{c}(M_\star,z)}^z \mathrm{d}z'\frac{\mathrm{d}t}{\mathrm{d}z'}\ 10^{\gamma(z', M_\star,z)} \frac{\msun}{\mathrm{yr}} \\
    & = M^c_\star(M_\star,z) + \sum_{j=N_c}^{N} 10^{\gamma(z_j, M_\star,z)} \frac{\msun}{\mathrm{yr}} \times (t(z_j) - t(z_{j-1}))  \nonumber
    \label{eq:intms}
\end{align}
Here $\gamma(z_j, M_\star,z) = -\alpha(M_\star, z)(1+z_j) + \beta(M_\star, z)$ is our fitted SFH; $z_\mathrm{c}$, $M^\mathrm{c}_\star$ and $N_\mathrm{c}$ are the critical redshift, mass and snapshot at which galaxies transition from stochastic to steady star formation, respectively; $N$ is the number of snapshots until and including $z$, and $t(z)$ the cosmic time at $z$.  We note that we apply the same redshift bins for the fitted SFH $\gamma(z_j,M_\star,z)$ as have been used for the {\sc astraeus} simulations, i.e. $z_j$ denotes the redshifts of the snapshots of the {\sc vsmdpl} simulation. Essentially, this equation states that the total stellar mass is a sum of that built up in the stochastic phase (first term on the RHS) and in the continuously star forming phase fit by a simple power-law (second term on the RHS). In Fig.~\ref{fig:msfitted} we validate our model by comparing the stellar mass we recover by using Eqn.~\ref{eq:intms} to the stellar mass directly obtained from the \textsc{astraeus} simulation for the \textit{Photoionization} model at $z=10-5$. To do so, we bin the galaxy sample in stellar mass ($M_\star$), compute their predicted stellar mass using our fitted SFH ($M_\star^\mathrm{fit}$) and take the median of $M_\star^\mathrm{fit}$ in each bin. This comparison is done down to $M_\mathrm{stoc}$, the minimum stellar mass at which more than 80\% of the stellar mass has been formed in a steady phase. To evaluate how robustly we recover the stellar mass, we introduce an uncertainty on the fitted SFH by drawing its normalization from a Gaussian distribution centered around $\beta(z, M_\star)$ (shown in Fig.~\ref{fig:beta}) with a standard deviation of $\sigma = 0.3$.

At all redshifts, we recover the stellar mass to an excellent degree for most of the mass range considered. The uncertainty in the normalization results in an uncertainty of $\sim 0.3\,\mathrm{dex}$ in $M_\star^\mathrm{fit}$. At the highest stellar masses, for example, above $M_\star \sim 10^{9.5}\,(10^{10.8})\,\mathrm{M_\odot}$ at $z=10\,(5)$, the predicted mass $M_\star^\mathrm{fit}$ oscillates around the one obtained from \textsc{astraeus}, as a result of the uncertainty in $\beta(z, M_\star)$ associated with the low number of galaxies at such high masses. This is a validation of the fact that observed stellar mass values can be successfully used to derive a SFH using our fits for $M_\star \sim 10^{7.5-9.8}\msun$ at $z \sim 10$ and $M_\star \sim 10^{8-10.5}\msun$ at $z \sim 5$.

\begin{figure*}
    \centering
    \includegraphics[width=\textwidth]{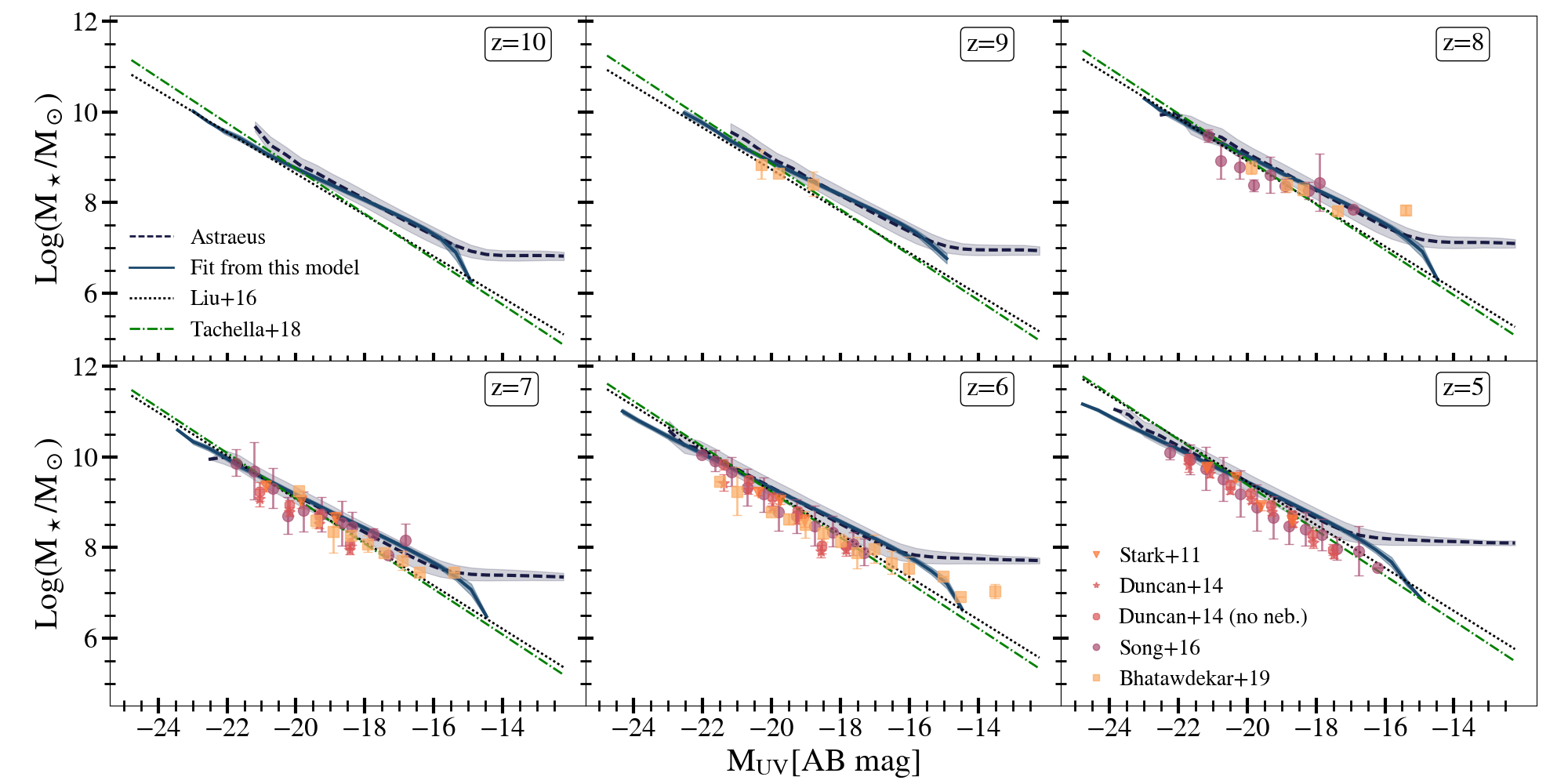}
    \caption{Median stellar mass ($M_\star$) as a function of the absolute UV magnitude directly obtained from the \textsc{astraeus} simulation (purple dashed line) and using our fitted SFH (blue solid line) for each galaxy in the \textit{Photoionization} model. The shaded purple area represents the standard deviation within \textsc{astraeus} and the shaded blue area is the standard deviation obtained by assuming a Gaussian spread ($\sigma = 0.3$) around the average value of $\beta(z, M_\star)$. We also plot observations from \citet{Stark_2011}, \citet{Duncan_2014}, \citet{Song_2016} and \citet{Bhatawdekar_2019}, and results from simulations from \citet{Liu2016} and \citet{Tacchella18}, as marked.}
    \label{fig:uvfitted}
\end{figure*}

Next, we check the mass-to-light ratios obtained from our fits as compared to those from \textsc{astraeus}; this is a crucial test of the model given that the UV luminosity is essentially dominated by star formation in the last few tens of Myrs. For each galaxy, we calculate its UV luminosity, $L_\mathrm{UV, tot}$, by convolving its fitted SFH, $10^{\gamma (z, \alpha, \beta)}~\msun/\mathrm{yr}$, with the UV luminosity evolution of a starburst, $\xi_\mathrm{SP}(t)$, \citep[see Eq.~16 in][]{Hutter2021astraeusI}. To model $\xi_\mathrm{SP}(t)$, we use the \textsc{starburst99} stellar population synthesis model assuming the previously specified Salpeter IMF and a metallicity of $Z=0.05 Z_\odot$.
\begin{align}
    L_\mathrm{UV,tot}(M_\star,z) & = M^c_\star(M_\star,z) \times \eta(t(z),t(z_c(M_\star,z))) \\
    & + \int_{z_c(M_\star,z)}^z \mathrm{d}z'\frac{\mathrm{d}t}{\mathrm{d}z'} 10^{\gamma(z',M_\star,z)}\frac{\msun}{\mathrm{yr}} \times \xi_\mathrm{SP}(t(z),t(z'))  \nonumber \\
    & = M^c_\star(M_\star,z) \times \eta(t(z),t(z_c(M_\star,z))) \nonumber\\
    & + \sum_{j=N_c}^{N} [ 10^{\gamma(z_j, M_\star,z)} \frac{\msun}{\mathrm{yr}} \times \eta(t(z),t(z_j)) \nonumber \\
    & \hspace{0.9cm} \times [t(z_j) - t(z_{j-1})] ] \nonumber
\end{align}
Analogous to the computation of the stellar mass from the fitted SFH, we apply the \textsc{vsmdpl} redshift bins to the fitted SFH and assume that the SFR within a redshift step remains constant. This reduces the integral to a sum, where we account for the constant star formation within a redshift step by introducing a correction factor $f_\mathrm{lin}$, and $\eta(t, t_j) = \xi_\mathrm{SP}(t, t_j) f_\mathrm{lin}(t, t_j, t_{j-1})$ \citep[see Eq.~14 and 15 in][]{Hutter2021astraeusI}.
The first term on the right-hand side is the UV luminosity from the stochastically formed stellar mass, while the second term on the right-hand side depicts the UV luminosity from the SFH part that is described by our fitting function. Assuming a Gaussian error with $\sigma=0.3$ for $\beta(M_\star, z)$, we find the scatter in the $M_\star-M_\mathrm{UV}$ relation to be less than $0.3$~dex (see Fig.~\ref{fig:uvfitted}).

Fig.~\ref{fig:uvfitted} shows the stellar mass as a function of the UV magnitude at $z=5-10$ for both galaxies simulated with \textsc{astraeus} and using our fitted SFH. Firstly, we note that the shown $M_\star$-$M_\mathrm{UV}$ relations includes only galaxies with $M_\star > M_\mathrm{stoc}$.\footnote{Due to their stochastic star formation, the star formation rates and hence UV luminosities of low-mass galaxies will greatly vary among galaxies with similar stellar masses. For this reason, applying a sharp cut in stellar or halo mass leads to a flattening of the $M_\star$-$M_\mathrm{UV}$ relation at low UV luminosities.} Secondly, our fitted SFHs yield a M$_\star$-M$_\mathrm{UV}$ relation in agreement, within uncertainties, to the \textsc{astraeus} results for $\muv \sim -15.5$ to $-20.5$ at $z \sim 10$ and $\muv \sim -17$ to $-23$ at $z \sim 5$. At the bright end ($\muv \lsim -20.5\, (-23)$ at $z \sim 10\, (5)$), however, we find the $M_\star-M_\mathrm{UV}$ relation directly inferred from \textsc{astraeus} to randomly over- or underpredict the one derived from our fitted SFHs. This is due to the low numbers of luminous galaxies ($<5$) that are not sufficient to reproduce the {\it average} trend that is obtained with our fitted SFH.
In order to put the difference between \textsc{astraeus} and our fitted SFH into perspective, we compare these relations to observations. Overall, we find our and the observational $M_\star$-$M_\mathrm{UV}$ relations to agree within their uncertainties. However, we see that we overpredict the stellar mass of fainter galaxies with $M_\mathrm{UV}\gtrsim-20$ at $z=5-6$ by about $0.2$~dex compared to the observations. This systematic deviation could be explained as follows: Firstly, the luminosities of our simulated galaxies do not include nebular emission. Its inclusion would shift the $M_\star$-$M_\mathrm{UV}$ relation towards higher UV luminosities for a given stellar mass. Nevertheless, when comparing the relations including (red circles) and not including nebular emission (red stars) from \citet{Duncan_2014}, we can see that the inclusion of nebular emission can shift the $M_\star$-$M_\mathrm{UV}$ relation to lower stellar masses by only $\sim0.1$~dex. Secondly, the stellar masses derived from the observed SEDs depend strongly on the assumed slope of the SFHs \citep[see Sec.~D in][]{Behroozi2019}. 
For instance, re-analysing the \citet{Song_2016} data with shallower SFHs i.e. with $\mathrm{SFR}\propto t^2$ instead of $\mathrm{SFR}\propto t^{5.5}$ as in \citet{Song_2016}, that better match the evolution of the cosmic star formation rate, UV LF and SMF evolution, increases the inferred stellar masses of the \citet{Song_2016} data points by $\sim0.2$~dex at all UV luminosities \citep{Behroozi2019}. This brings them into perfect agreement with our simulation results. 
Thus, we note that the uncertainties introduced by our SFH fitting function (Eqn. \ref{eq:fit}) are negligible compared to the observational uncertainties.

Finally, we compare our $M_\star$-$M_\mathrm{UV}$ relations to those obtained with the \textsc{meraxes} \citep[black dotted line;][]{Liu2016} and Tacchella \citep[green dash-dotted line;][]{Tacchella18} semi-analytic models. 
Firstly, while the $M_\star$-$M_\mathrm{UV}$ relations of these models agree with those obtained from \textsc{astraeus} for bright galaxies ($M_\mathrm{UV}\lesssim-20$), they show a steeper slope and thus lower stellar masses for UV faint galaxies ($M_\mathrm{UV}\gtrsim-20$). The reason for this difference can be explained as follows: the  \textsc{astraeus} simulations do not include dust or account for dust attenuation but adjust the limiting star formation efficiency $f_\star$ to the dust attenuated UV luminosity functions (UV LFs). However, since massive and bright galaxies experience stronger dust attenuation than low-mass and faint galaxies, the slopes of the simulated UV LFs and $M_\star$-$M_\mathrm{UV}$ relations become shallower as when accounting for dust attenuation. 
Secondly, we note that the difference between the normalization offsets of the \textsc{astraeus} and \textsc{meraxes} or Tacchella $M_\star$-$M_\mathrm{UV}$ relations decrease with decreasing redshift; albeit this difference remains small. While both the \textsc{meraxes} and Tacchella models basically assume redshift-independent star formation efficiencies, the star formation efficiency in \textsc{astraeus} is redshift dependent and decreases for low-mass galaxies with decreasing redshift (see Fig. \ref{fig:feffMgasIni}). The fact that the difference of the normalization offsets is smallest at the lowest redshifts shown is then expected, as \citet{Tacchella18} derive their model star formation efficiencies by calibrating to the $z=4$ UV LF.

\section{Conclusions and discussion} \label{Conclusions}

In this work, we have used the \textsc{astraeus} (semi-numerical rAdiative  tranSfer coupling of galaxy formaTion and Reionization in N-body dArk mattEr simUlationS) framework, which couples an N-body simulation with a semi-analytical galaxy formation model and a semi-numerical reionization scheme. Our aim is to quantify the star formation histories of galaxies during the EoR for different radiative feedback models, ranging from weak and delayed feedback to strong and instantaneous feedback. 

We find the star formation in low-mass galaxies ($M_\mathrm{h}\lesssim10^{9.3}\msun$) to be stochastic (stars form at a rate that deviates from the SFH fit described by Eq.~\ref{eq:fit} by more than $\Delta_\mathrm{SFR}=0.6\,\mathrm{dex}$), and to transition to continuous as galaxies become more massive and less governed by SN and radiative feedback. In order to describe the SFH of a given galaxy, we have investigated in a first step how the fraction of stellar mass formed during the initial phase of stochastic star formation evolves with redshift, stellar mass, and depends on the assumed radiative feedback model. In a second step, we have fit the SFHs of galaxies at $z\geq5$ with a power law such that
\begin{eqnarray*}
    \mathrm{\log_{10}}\left(\frac{\mathrm{SFR}(z)}{\msun/\mathrm{yr}}\right) &=& - \alpha(M_\star, z_\mathrm{obs})(1+z) + \beta( M_\star, z_\mathrm{obs}).
\end{eqnarray*}
Our four radiative feedback models comprise scenarios where the gas mass that a galaxy in an ionized region can maintain is given by the filtering mass \citep{Gnedin2000, Naoz_2013} that is determined by the temperature of the photo-heated gas or the photoionization rate when the region becomes reionized, or the Jeans mass at the virial over-density.
Our main results are: 
\begin{enumerate}
    \item At every redshift, the fraction of stellar mass formed stochastically $M_\star^\mathrm{c}/M_\star$ and the fraction of time spent in the stochastic phase $t^\mathrm{stoc}/t^\mathrm{tot}$ decrease with increasing stellar mass $M_\star$ and decreasing redshift for galaxies with $M_\star < 10^{8.5}\,\mathrm{M_\odot}$.  These quantities hardly vary for the different time delayed radiative feedback models. Only for the instantaneous strong radiative feedback model, the \textit{Jeans Mass} model,  we find $M_\star^\mathrm{c}/M_\star$ and $t^\mathrm{stoc}/t^\mathrm{tot}$ to show higher values for low-mass galaxies with $M_\star < 10^{8.5}\,\mathrm{M_\odot}$.
    \item For galaxies with $M_\star \gtrsim 10^{8.5}\,\mathrm{M_\odot}$ the SFH increases continuously with time following the power law specified in Eqn. \ref{eq:fit}. Its slope $\alpha$ scales with the effective star formation efficiency of a galaxy. The lower the galaxy's stellar mass is, the stronger is on average the suppression of star formation by SN and radiative feedback and the lower the average $\alpha$ value for the delayed radiative feedback models. However, for the strong instantaneous radiative feedback model, we find the average $\alpha$ values to increase towards lower stellar masses due to galaxies more affected by radiative feedback being removed from the sample. As the star formation in a galaxy becomes less affected by the feedback processes, its SFH slope $\alpha$ converges to a constant value of $\sim 0.18$. The stellar mass $M_\mathrm{stoc}$, at which this transition from rising to constant occurs, increases as the galaxy's gravitational potential becomes shallower with decreasing redshift, going from $\sim 10^8\,\mathrm{M_\odot}$ at $z=10$ to $\sim 10^9\,\mathrm{M_\odot}$ at $z=5$. 
    \item Given that the SFH slopes $\alpha$ converge to a single value for massive galaxies ($M_\star>M_\mathrm{stoc}$), the corresponding normalization $\beta$ of the power law describing the SFH increases with rising stellar mass, going from $\sim0$ for galaxies with $M_\star = 10^{8.2}\msun$ to $\sim3.5$ for galaxies with $M_\star = 10^{11}\msun$ at $z=5$. $\beta$ also increases with increasing redshift, e.g increasing for galaxies with $M_\star = 10^9\msun$ from $\beta=1$ at $z=5$ to $\beta = 3$ at $z=10$. For low-mass galaxies ($M_\star \sim M_\mathrm{stoc}$), the normalization is strongly correlated with the SFH slope $\alpha$, reflecting that the same stellar mass can be accumulated either over a long time with a low SFR or over a shorter time with a higher SFR.
    \item For each radiative feedback model, we provide the fitting function to the continuously rising part of our simulated SFHs. Integrating these fitting functions over time and accounting for the stellar mass accumulated in the stochastic phase at the beginning, we recover the stellar masses of all simulated galaxies within an uncertainty of $0.1\,$dex for $M_\star \sim 10^{7.5-9.8}\msun$ at $z \sim 10$ and $M_\star \sim 10^{8-10.5}\msun$ at $z \sim 5$. Our fitted SFHs yield a M$_\star$-M$_\mathrm{UV}$ relation in agreement, within uncertainties, to the \textsc{astraeus} results for $\muv \sim -15.5$ to $-20.5$ at $z \sim 10$ and $\muv \sim -17$ to $-23$ at $z \sim 5$.
\end{enumerate}

There are a few caveats to the work presented in this paper. 
Firstly, as mentioned in Sec.~\ref{characterizing} and Appendix \ref{app:cuts} and \ref{app:stoc}, the stellar mass formed and time spent in the stochastic phase, $M_\star^\mathrm{c}/M_\star$ and $t^\mathrm{stoc}/t^\mathrm{tot}$, depend on the stochasticity and selection criteria. Nevertheless, we note that the found trends and values of the SFH slope $\alpha(M_\star, z)$ and normalization $\beta(M_\star, z)$ are robust and are not dependent on the stochasticity criteria.
Secondly, as we mention in Sec.~\ref{properties}, we define the lifetime of a galaxy as the duration of its mass assembly, which depends on the mass resolution of the underlying merger trees. Assuming a mass-weighted lifetime would yield more robust results when changing the underlying mass resolution of the merger trees and/or N-body simulation, and shift the time spent in the stochastic phase, $t^\mathrm{stoc}/t^\mathrm{tot}$, to lower values as galaxies with higher stellar masses are considered. 
Thirdly, the functional form of our SFH fitting function can not reproduce the flattening of the SFHs (and DM assembly histories) that we see for the massive galaxies in our simulation. Hence, we overpredict the recent SFRs and hence UV luminosities of bright galaxies i.e. those with $\muv \lsim -20.5\, (-23)$ at $z \sim 10\, (5)$ in Fig. \ref{fig:uvfitted}.
Lastly, the \textsc{astraeus} model used in this work does not account for dust attenuation of the UV. Including a description for dust and its attenuation of the UV would predominantly affect the properties of massive galaxies. The stellar masses for a given UV luminosity would increase and steepen the stellar mass - UV luminosity relation in Fig. \ref{fig:uvfitted} at the bright end. We will assess the effects of dusts in a forthcoming \textsc{astraeus} version that incorporates a dust model.

With the launch of JWST, NGRST and Euclid, the number of observed high-redshift galaxies during the EoR will increase greatly. The NGRST High Latitude Survey (HLS) and Euclid will probe only the brightest galaxies that are well into the phase of continuous star formation: while  Euclid will integrate down to and $M_\mathrm{UV} \leq -21\,(-23)$ at $z=5\,(10)$\footnote{https://sci.esa.int/web/euclid/-/euclid-nisp-instrument}, the HLS survey has a UV magnitude limit of $M_\mathrm{UV} \leq -20.4$ and $-20.6$ at $z=8$ and 10, respectively \citep{Waters2016}. Further, JWST surveys, such as JADES, will have the potential to observe even galaxies that form stars stochastically with limits of $\muv \leq -16$ and $-18$ at $z=5$ and 10, respectively \citep{Williams2018}. As shown, our model, that can recover the M$_\star$-M$_\mathrm{UV}$ relation and SFHs for galaxies with $\muv \sim -15.5$ to $-20.5$ at $z \sim 10$ and $\muv \sim -17$ to $-23$ at $z \sim 5$, will be extremely useful in shedding light on the assembly histories of the galaxies observed by these forthcoming facilities.

\section*{Acknowledgements}

The authors thank the anonymous referee for their comments that improved the quality of the paper. All authors acknowledge support from the European Research Council's starting grant ERC StG-717001 (``DELPHI"). PD acknowledges support from the NWO grant 016.VIDI.189.162 (``ODIN") and the European Commission's and University of Groningen's CO-FUND Rosalind Franklin program.
GY acknowledges financial support from MINECO/FEDER under project grant AYA2015-63810-P and MICIU/FEDER under project grant  PGC2018-094975-C21.
The authors wish to thank V. Springel for allowing us to use the L-Gadget2 code to run the different Multidark simulation boxes, including the \textsc{vsmdpl} and \textsc{esmdpl} used in this work. The \textsc{vsmdpl} and \textsc{esmdpl} simulations have been performed at LRZ Munich within the project pr87yi. The CosmoSim database (\url{www.cosmosim.org}) provides access to the simulation and the Rockstar data. The database is a service by the Leibniz Institute for Astrophysics Potsdam (AIP)

\section*{Data Availability} 

The source code of the semi-numerical galaxy evolution and reionization model within the \textsc{astraeus} framework and the employed analysis scripts are available on GitHub (\url{https://github.com/annehutter/astraeus}). The underlying N-body DM simulation, the \textsc{astraeus} simulations and derived data in this research will be shared on reasonable request to the corresponding author.




\bibliographystyle{mnras}
\bibliography{mybib} 





\appendix

\section{Galaxy sample cuts} \label{app:cuts}

\begin{figure}
\centering
\begin{subfigure}[b]{\columnwidth}
   \includegraphics[width=1\linewidth]{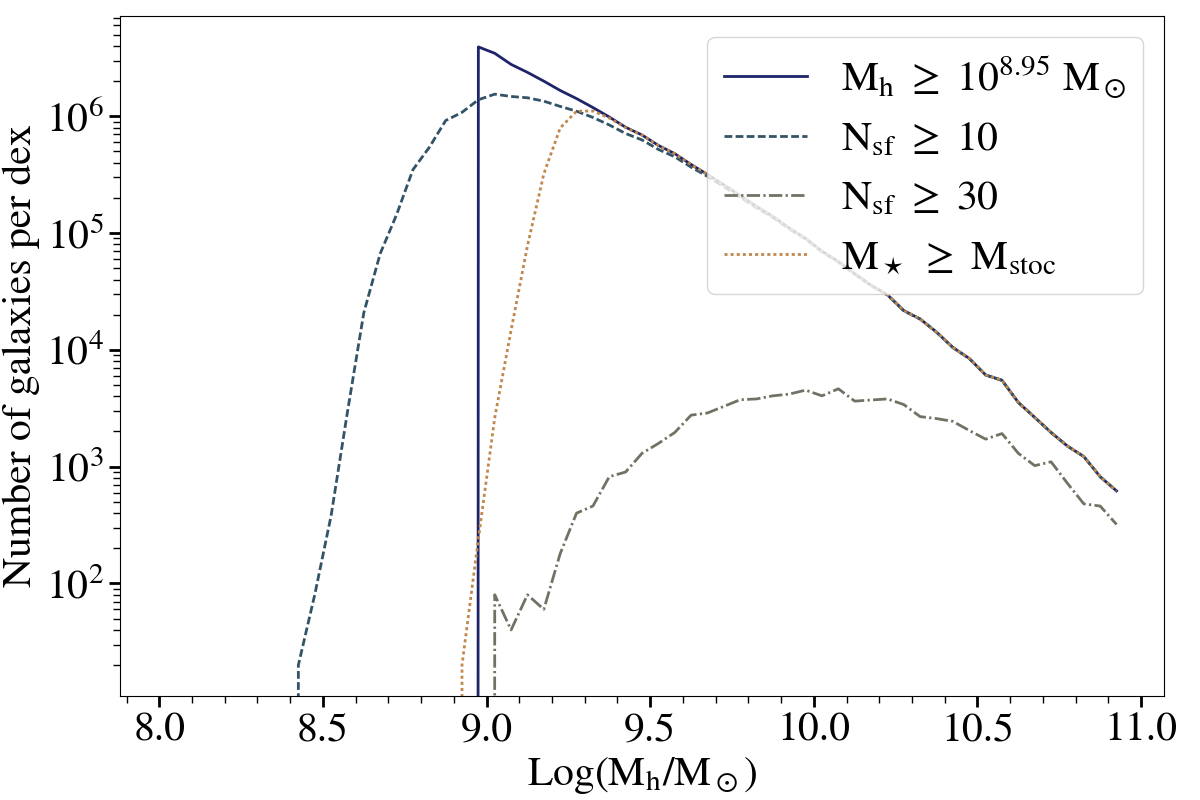}
\end{subfigure}

\begin{subfigure}[b]{\columnwidth}
   \includegraphics[width=1\linewidth]{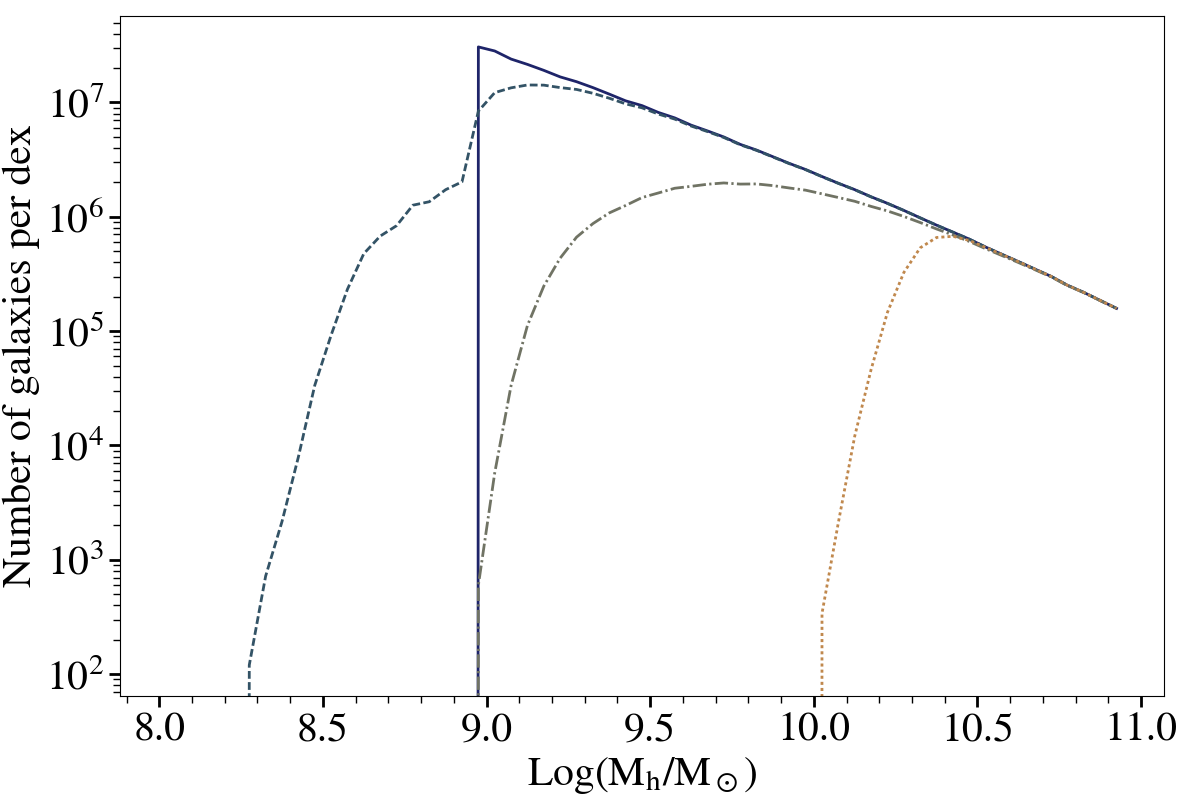}
\end{subfigure}

\caption{Number of galaxies as a function of halo mass for different selection criteria at $z=10$ (top panel) and at $z=5$ (bottom panel)  in the \textit{Photoionization} model.}
\label{fig:cuts}
\end{figure}

While we have applied two selection cuts to the galaxy sample used for the results in Sec.~\ref{Results} - considering only galaxies with converged SFHs ($M_h\geq10^{8.95}\msun$) \textit{and} with star formation in at least the last $10$ consecutive redshift steps ($N_\mathrm{SF}\geq10$) -, we comment in this Section on how our results, in particular the stellar mass accumulated in the stochastic phase ($M_\star^\mathrm{c}$), change as we relax the convergence criterion or alter the necessary number of consecutive redshift steps with star formation $N_\mathrm{SF}$. We show how the different selection cuts alter the number of galaxies as a function of halo mass $M_h$ at $z=10$ and $z=5$ in Fig. \ref{fig:cuts}, respectively. Relaxing the convergence criterion, i.e. including also galaxies with $M_h<10^{8.95}\msun$ and $N_\mathrm{SF}\geq10$ (c.f. dashed lines), enhances the number of low-mass galaxies ($M_\star\lesssim10^{6.5}\msun$) with continuous star formation and leads to a decrease of the stellar mass formed during the phase of stochastic star formation, lowering the corresponding  $M_\star^\mathrm{c}/M_\star$ and $t^\mathrm{stoc}/t^\mathrm{tot}$ and $M_\star^\mathrm{c}$ values while enhancing the $z_c$ values. In contrast, increasing $N_\mathrm{SF}$ removes short-lived galaxies, which correspond to removing increasingly lower mass galaxies with decreasing redshift (c.f. dash-dotted lines): $M_\star^\mathrm{c}/M_\star$ and $t^\mathrm{stoc}/t^\mathrm{tot}$ decreases towards lower mass galaxies. From Fig. \ref{fig:cuts}, we also note that selecting galaxies with $N_\mathrm{SF}\geq30$ would result in selecting only galaxies in the phase of continuous star formation at $z=10$, while would include galaxies in the stochastic phase at $z=5$ (c.f. dash-dotted lines).

\section{Determining the stochasticity criterion} \label{app:stoc}

\begin{figure}
\centering
\begin{subfigure}[b]{\columnwidth}
   \includegraphics[width=1\linewidth]{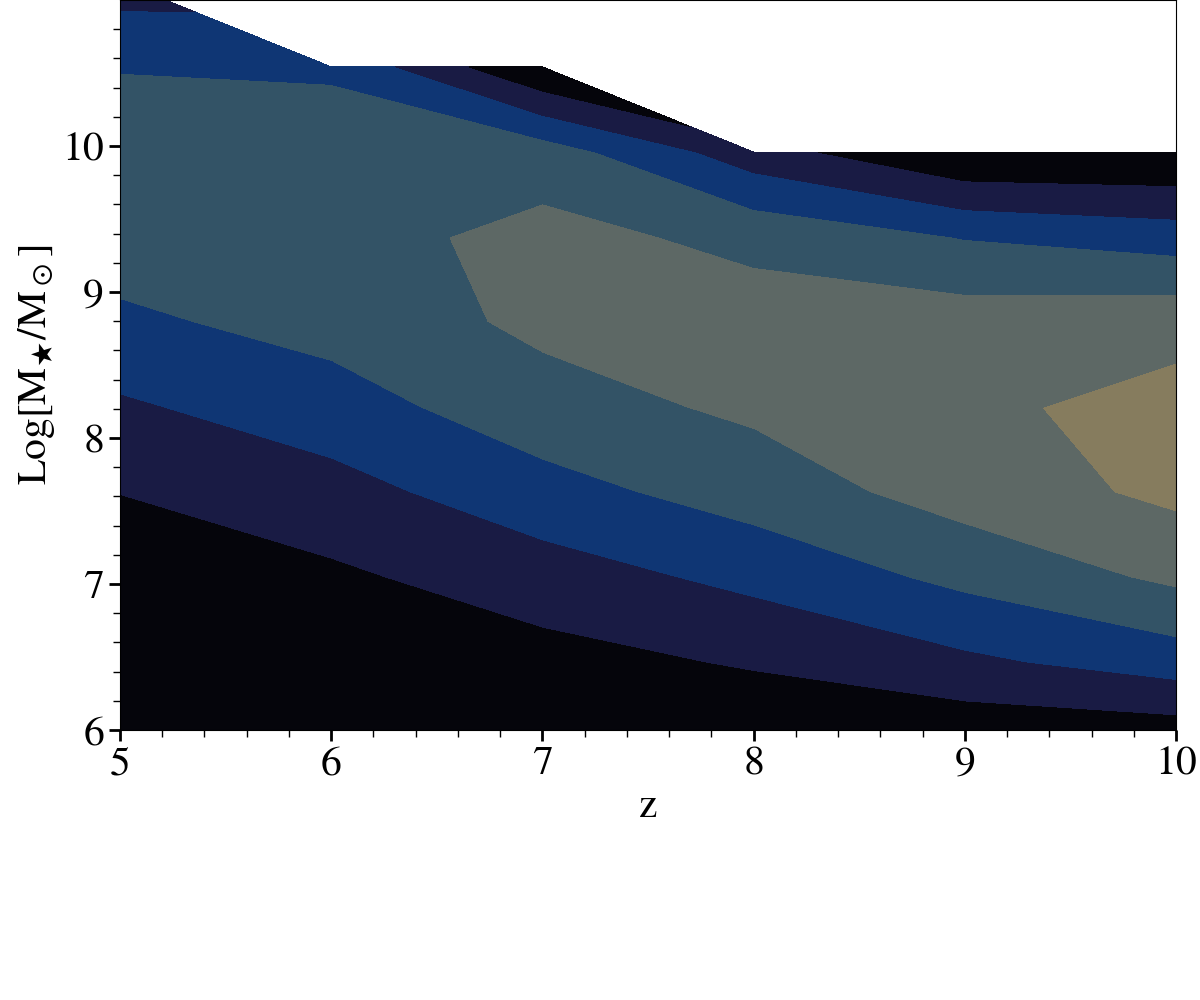}
\end{subfigure}

\begin{subfigure}[b]{\columnwidth}
   \includegraphics[width=1\linewidth]{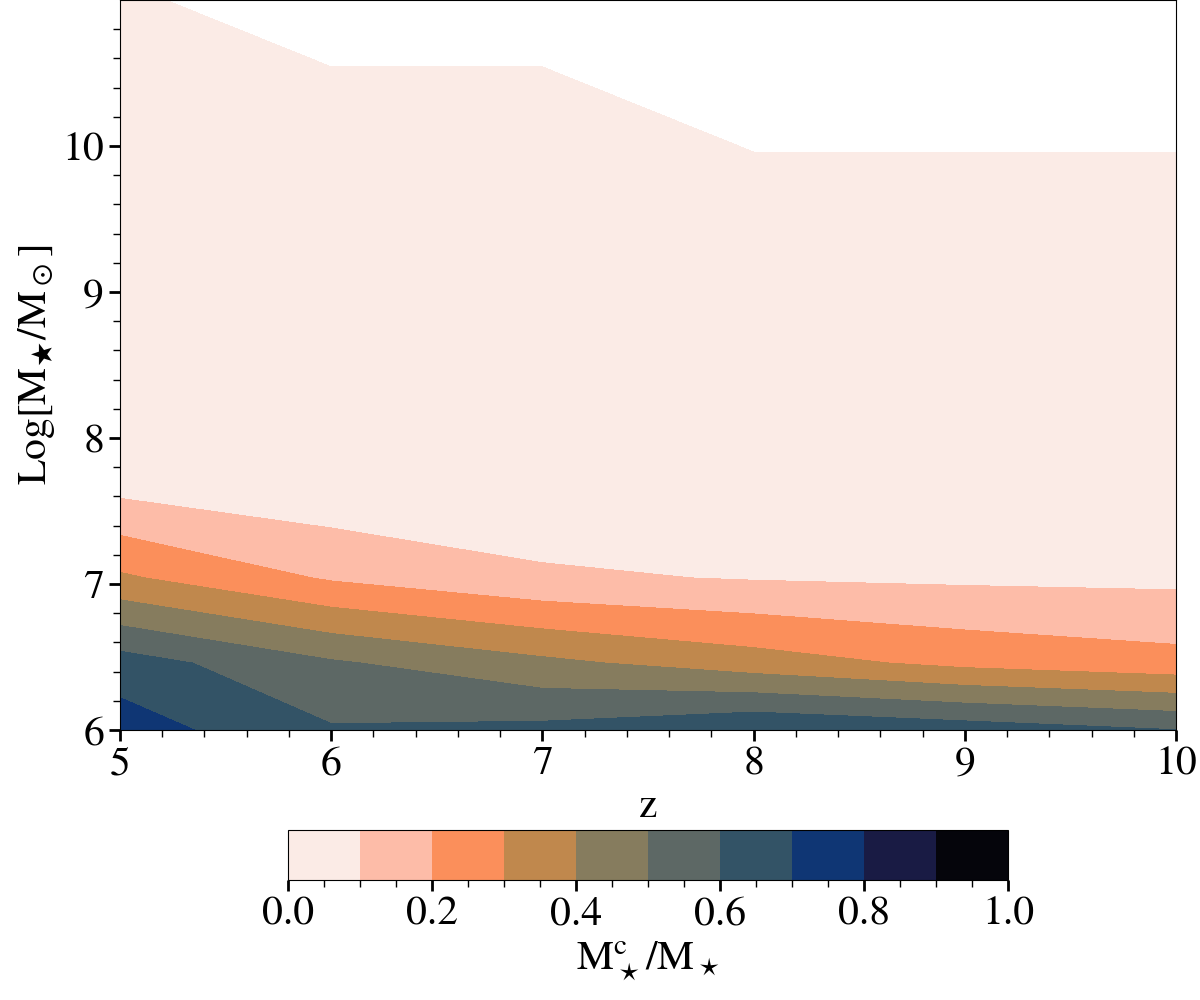}
\end{subfigure}

\caption{Mean fraction of stellar mass formed in the stochastic phase as a function of redshift and stellar mass for \textit{Photoionization} model, assuming that a galaxy is stochastic if its instantaneous SFR deviates by more than $0.2\,\mathrm{dex}$ (top panel) or $1\,\mathrm{dex}$ (bottom panel) compared to the linear regression of its SFH.}
\label{fig:delta}
\end{figure}

As outlined in Sec.~\ref{characterizing}, the key criterion that defines whether a galaxy forms stars stochastically is the deviation from the linear regression from its SFH, $\Delta_\mathrm{SFR}$. Here we briefly discuss how our results change as the criterion $\Delta_\mathrm{SFR}=0.6$ assumed throughout the paper is altered. 
As the $\Delta_\mathrm{SFR}$ value is increased, galaxies that would have been identified as being in the stochastic phase before are classified then as galaxies with continuous star formation. As a consequence, less stellar mass is formed in the stochastic phase, which we can see when we compare the $M_\star^\mathrm{c}/M_\star$ values for $\Delta_\mathrm{SFR}=1$ in the bottom panel of Fig. \ref{fig:delta} with those for $\Delta_\mathrm{SFR}=0.6$ in Fig. \ref{fig:sto}. However, the trends of $M_\star^\mathrm{c}/M_\star$ with stellar mass and redshift persist.
In contrast, as the $\Delta_\mathrm{SFR}$ value is reduced, we find a higher fraction of stellar mass being formed stochastically as can be seen when comparing the top panel of Fig.~\ref{fig:delta} for $\Delta_\mathrm{SFR}=0.2$ with Fig. \ref{fig:sto} for $\Delta_\mathrm{SFR}=0.6$. Interestingly, we also find that $M_\star^\mathrm{c}/M_\star$ increases as we go from intermediate massive ($M_\star \sim 10^{10}\,(10^{8.5})\,\mathrm{M_\odot}$ at $z=5\,(10)$) to the most massive galaxies ($M_\star \geq 10^{10.8}\,(10^{9.8})\,\mathrm{M_\odot}$ at $z=5\,(10)$). This trend traces back to the DM assembly histories of the massive galaxies that shape the corresponding SFHs. Since the flattening of their slopes towards lower redshifts is not captured by our SFH fitting function, the corresponding shallower SFHs lie then outside the $\Delta_\mathrm{SFR}$ margin and are marked as forming stars stochastically.

\section{SFH fitting parameters} \label{Appendix1}

In this Section we present the fitting parameters $\alpha$ and $\beta$ in Eqn. \ref{eq:fit} for all stellar masses $M_\star$, redshifts $z$ and radiative feedback models covered in this work.
For the \textit{Photoionization}, \textit{Early Heating} and \textit{Strong Heating} models, we fit $\alpha(z, M_\star)$ with the following fitting function
\begin{multline}
    \alpha(z, M_\star) = (a_\alpha \times z+b_\alpha) \mathrm{exp}\bigg(-10^{\frac{c_\alpha}{(z-d_\alpha)^{e_\alpha}}-\mathrm{Log}(M_\star)}\bigg),
    \label{eq:alpha1}
\end{multline} 
while for the \textit{Jeans Mass} model, we use
\begin{multline}
    \alpha(z, M_\star) = (a_\alpha \times z+b_\alpha) \mathrm{exp}\bigg(10^{c_\alpha \times z + d_\alpha-\mathrm{Log}(M_\star)}\bigg).
    \label{eq:alpha2}
\end{multline} 
For each model, we show the values of all free parameters present in Eqn.~\ref{eq:alpha1} and Eqn.~\ref{eq:alpha2} in Table \ref{tab:alpha}.

\begin{center}
\begin{table}
\caption{\label{tab:alpha} For the radiative feedback model shown in column 1, we show the value taken by the parameters in Eq.~\ref{eq:alpha1} for the first three models and in Eq.~\ref{eq:alpha2} for the \textit{Jeans Mass} model.}
\begin{tabular}{ |c|c c c c c| } 
 \hline
 Models & a$_\alpha$ & b$_\alpha$ & c$_\alpha$ & d$_\alpha$ & e$_\alpha$ \\
 \hline
 Phoionization & 0.0025 & 0.1661 & 6.9344 & 4.7825 & 0.0366 \\ 
 \hline
 Early Heating & 0.0024 & 0.1662 & 6.9312 & 4.7931 & 0.03517 \\
 \hline
 Strong Heating & 0.0023 & 0.1665 & 7.0920 & 4.6371 & 0.0467 \\
 \hline
 Jeans Mass & 0.0054 & 0.1669 & -0.4117 & 9.8585 & X \\
 \hline
\end{tabular}
\end{table}
\end{center}

For the \textit{Photoionization}, \textit{Early Heating} and \textit{Strong Heating} models, we fit  $\beta(z, M_\star)$ with the following fitting function
\begin{multline}
   \beta(z, M_\star) = (a_\beta \times z+b_\beta)\mathrm{Log}(M_\star)^2 + (c_\beta \times z + d_\beta)*\mathrm{Log}(M_\star) + \\ e_\beta \times z + f_\beta,
   \label{eq:beta1}
\end{multline}
while for the \textit{Jeans Mass} model, we use
\begin{eqnarray}
   \beta(z, M_\star) &=& (a_\beta \times z+b_\beta)\mathrm{Log}(M_\star) + c_\beta \times z + d_\beta.
   \label{eq:beta2}
\end{eqnarray}
For each model, we show the values of all free parameters present in Eqn.~\ref{eq:beta1} and Eqn.~\ref{eq:beta2} in Table \ref{tab:beta}.

\begin{center}
\begin{table}
\caption{\label{tab:beta} For the radiative feedback model shown in column 1, we show the value taken by the parameters in Eq.~\ref{eq:beta1} for the first three models and in Eq.~\ref{eq:beta2} for the \textit{Jeans Mass} model.}
\begin{tabular}{ |c|c c c c c c| } 
 \hline
  Models & $a_\beta$ & $b_\beta$ & $c_\beta$ & $d_\beta$ & $e_\beta$ & $f_\beta$\\
 \hline
 Phoionization & -0.028 & 0.126 & 0.481 & -1.008 & -1.720 & -1.358 \\ 
 \hline
 Early Heating & -0.028 & 0.125 & 0.478 & -1.012 & -1.710 & -1.313 \\
 \hline
 Strong Heating & -0.029 & 0.129 & 0.501 & -1.040 & -1.790 & -1.400 \\
 \hline
 Jeans Mass & -0.031 & 0.943 & 0.364 & -8.499 & X & X \\
 \hline
\end{tabular}
\end{table}
\end{center}


\bsp	
\label{lastpage}
\end{document}